\documentclass[aps,pre,twocolumn,amssymb,floatfix,showpacs]{revtex4}
\usepackage{epsfig}

\newcommand{\Maybeeq}[1]{Eq.~(#1)}
\newcommand{\Maybeeqs}[1]{Eqs.~(#1)}
\newcommand{\maybeeq}[1]{Eq.~(#1)}
\newcommand{\maybeeqs}[1]{Eqs.~(#1)}

\newcommand{\cc}{{\rm c}}
\newcommand{\op}{{\rm op}}
\newcommand{\loc}{{\rm loc}}
\newcommand{\lcftbdy}{\partial\theta\partial\bar\theta}
\newcommand{\lcftbulk}{\partial\theta\bar\partial\bar\theta+
\bar\partial\theta\partial\bar\theta}
\newcommand{\tree}{\rm Tree}
\newcommand{\NNP}{\rm NNP}
\newcommand{\N}{\mathcal{N}}
\newcommand{\id}{\mathbb I}
\newcommand{\vi}{\vec{i}}
\newcommand{\vj}{\vec{j}}
\newcommand{\htpair}[2]{(h_{\vi},h_{\vj})=(#1,#2)}
\newcommand{\DelSubs}{\Delta_{\vi\;\vj}}
\newcommand{\DelSubsii}{\Delta_{\vi\;\vi}}

\begin{document}

\title{The four height variables, boundary correlations, 
and dissipative defects
in the Abelian sandpile model}

\author{M. Jeng}

\email{mjeng@siue.edu}

\affiliation{
Box 1654, Department of Physics, Southern Illinois
University Edwardsville, Edwardsville, IL, 62025}


\begin{abstract}

\noindent We analyze the two-dimensional Abelian
sandpile model, and demonstrate that the four height
variables have different field identifications
in the bulk, and along closed boundaries, but become
identical, up to rescaling, along open boundaries. We
consider
two-point boundary correlations in
detail,
and discuss a number of complications that arise in the 
mapping from sandpile correlations to spanning tree
correlations;
the structure of our results suggests a conjecture that
could greatly simplify future calculations.
We find a number of three-point functions
along closed boundaries, and propose closed boundary
field identifications for the height variables.
We analyze the effects of 
dissipative defect sites, at which the number
of grains is not conserved,
and show that dissipative defects along closed
boundaries, and in the bulk, have no effect on any 
weakly allowed cluster variables, or on their
correlations.
Along open boundaries, 
we find a particularly simple field structure;
we calculate all $n$-point
correlations, for any combinations of height
variables and dissipative defect sites, and find that 
all heights and defects are
represented by the same field operator.
\end{abstract}


\maketitle


\section{INTRODUCTION}

The Abelian sandpile model (ASM), 
introduced by Bak, Tang, and Wiesenfeld,
is the original prototype for 
self-organized criticality~\cite{BTW}. Systems
with self-organized criticality  are 
naturally driven to a critical point, and thus 
can  potentially 
explain how power laws occur in nature
without any fine-tuning of parameters.
Since their introduction, sandpile models have been used to 
model an extraordinarily wide range of systems, 
from earthquakes~\cite{earthquakes} to
river networks~\cite{rivers, rivers2}; 
see~\cite{BakBook,JensenBook} for reviews.

To be precise, we are considering the
two-dimensional isotropic Abelian sandpile model. This
is a very simple model; in fact, its simplicity is
its strength, since otherwise it could not act as a model
for such a diverse range of physical systems. The ASM is
defined on a square lattice, where each site has a height
variable (the number of grains of sand
at that site) that can
range from 1 to 4. At each time step, a grain of
sand is added to a random site. Any site with more than 
four grains is unstable, and collapses, losing four grains,
and sending one grain to each of its neighbors. Unstable
sites are repeatedly collapsed until all sites are
stable. Then, a new time step begins---a 
grain is added to a random site,
and the process 
begins anew~\cite{BTW}.
Initially, probabilities of configurations will
depend on the initial conditions, but after a long
period of time, the ASM develops a well-defined
probability distribution of states, independent of
the initial conditions~\cite{Dhar.First}. Typically,
the number of grains is conserved in each toppling,
except for sites along open boundaries, where grains
are lost with each toppling (i.e. fall off the
edge). There must be at least one dissipative
site---i.e. at least one site where
the number of grains decreases
upon toppling---or else
the sandpile would eventually reach a state
where topplings continued endlessly
during a single time step.

Despite its simplicity, certain basic properties
 of the ASM remain unknown.
For example, despite intensive
work, the power law governing the sizes of avalanches 
in the ASM is still 
unknown---see~\cite{DharReview} for a review. 
And while the 
height one variable is well understood, the roles
played by the higher height variables (two, three, and four)
are not. For example, 
no bulk two-point correlation
functions of higher height variables are 
known.

It is known that the ASM is related to the set of 
spanning trees that can be drawn on the sandpile
lattice, and that this relationship can be used to
perform exact calculations of ASM
probabilities~\cite{Dhar.First,Dhar.CFT}. A spanning 
tree
is a set of arrows drawn on the lattice, 
such that 
each site has exactly one arrow pointing 
from the site to a 
neighbor, and such that there are no closed loops of arrows.
Following the path of 
arrows from any site will eventually lead
off the edge of the sandpile (or, more generally,
to a
dissipative site, such as found on an open 
boundary)---the ``site''
off the edge of the sandpile is called the root.
A number of relationships between the
ASM and spanning tree states are known. For example,
the number of recurrent states of the ASM (states
that occur with nonzero probability after a long
amount of time) is equal to the number of
spanning trees that can be drawn on the sandpile
lattice~\cite{Dhar.CFT}. 

Spanning trees are, in turn, related to the $c=-2$
conformal field theory (CFT). The $c=-2$
CFT is the simplest known example of a logarithmic
conformal field theory (LCFT), and is well 
understood~\cite{c2Describe.1,c2Describe.2,c2Describe.3}. 

A method
introduced by Majumdar and
Dhar exploits the
mapping between ASM states and
spanning tree states to obtain exact ASM 
probabilities~\cite{Dhar.UnitCorrelations}.
It has long been known that the Majumdar-Dhar
method can be used to find
the two-point correlation
function of the unit height variable, which
decays as $1/r^4$~\cite{Dhar.UnitCorrelations}.
More recently, Mahieu and Ruelle used
the Majumdar-Dhar method to calculate correlation
functions of a number of ASM height configurations,
known as weakly allowed cluster
variables~\cite{Mahieu.Ruelle}.
They not only found that all the correlations decayed
as $1/r^4$, but were able to 
use their correlations to identify
the thirteen simplest weakly allowed cluster variables with 
operators in the LCFT.  These variables were all identified 
with linear combinations
of three LCFT  field operators,
all of which had scaling dimension two, but only
one of which---the $\lcftbulk$ operator---was isotropic.
In some ways, this suggested that
the higher height variables should
be identified with $\lcftbulk$; on
the other hand, Mahieu and Ruelle pointed out
that this appeared inconsistent with
LCFT operator product expansions (OPE's).

Despite the power of these mappings, and of the Majumdar-Dhar
method, fundamental questions about the ASM remain
unanswered, because  aspects of the mapping between
the ASM and $c=-2$ LCFT are still unknown---for example,
it is not known what field operators in the $c=-2$
LCFT  represent the higher height variables 
of the ASM (or, indeed, whether such a representation
even exists).
A single site with height two, or any higher height
variable, is not a weakly allowed cluster, and
thus higher height probabilities and correlations
cannot be calculated with the Majumdar-Dhar method.
Priezzhev was able to extend the Majumdar-Dhar
method to calculate the bulk probabilites for all higher
height variables~\cite{Priezzhev}. However, 
the bulk correlations of the higher height variables,
which would be needed to obtain the 
field identifications of the higher height
variables, remain unknown.

Ivashkevich calculated all two-point 
correlation functions
of all height variables, along open and closed
boundaries~\cite{Ivashkevich}. He found that
all boundary correlations, between all height variables,
decayed as $1/r^4$, and argued that this implied
that all four height variables should be represented
by the same field operator (up to rescaling). 
Dhar has argued that,
based on clustering properties of correlation
functions, the bulk correlations
should be expected to factorize in a manner consistent
with giving all four height variables the same
field identification~\cite{Dhar.AllSame}.

However, we argue here that the four height variables 
should in fact receive different field identifications,
both along closed boundaries, and in the bulk,
and propose field identifications along
closed boundaries.
Our conclusions are based on 
analysis of closed boundary
three-point functions, and 
of dissipative defect sites, as well as
a reanalysis of the methods and
results of Ivashkevich.
However, we show that
along open boundaries
all four height
variables, as well as dissipative defect sites,
are represented by the same operator,
$\lcftbdy$, in the $c=-2$ LCFT. 
We demonstrate this by computing all $n$-point
correlations of height variables
and dissipative defects.

In sections~\ref{sec:Green.Formalism} 
and~\ref{sec:Height.Probabilities}
we briefly review the methods used by Majumdar, Dhar,
and Priezzhev for studying the ASM.
In section~\ref{sec:Ivashkevich.Review}, we
review Ivashkevich's calculations of
the boundary height probabilities.

In section~\ref{sec:Closed.Loop.Calculation}, and
appendices~\ref{sec:Complication1}
and~\ref{sec:Complication2}, we discuss issues 
associated with boundary correlation functions.
While Ivashkevich has already calculated the 
boundary two-point correlations~\cite{Ivashkevich},
we show that he mischaracterized the 
mapping between ASM configurations and spanning tree
configurations, and a correct characterization results in a
number of complications, necessitating a reanalysis of the
two-point correlations. 
The relationship between ASM states and spanning tree
states is not what one might have initially expected;
we also note that linear relationships between nonlocal spanning
tree conditions and local spanning tree conditions 
for one-point probabilities
do not carry over in a simple fashion for multipoint
correlation functions. Both of these 
complications introduce what we call
``anomalous graphs''---while these complications are
important, because they are technical in nature, we
delegate much of the discussion to the appendices.
In section~\ref{sec:Closed.Loop.Calculation}
we calculate the anomalous graphs, and 
conjecture that the anomalous graphs have no
effect on the universal parts of any boundary
correlation functions; while we have not been able to 
prove this conjecture, it holds true for all correlation
functions that we have calculated.

In section~\ref{sec:Three.Point.Closed}, 
we look at correlation functions along closed boundaries.
For two-point correlation functions, we find that while
we disagree with Ivashkevich's relationship between
ASM and spanning tree states, we agree with his
final results. However, we argue that these final results
are, in fact, not consistent with identifying 
all height variables with the same field operator.
Next, we calculate all three-point functions along closed
boundaries that involve at least one unit height
variable, and use these to make field identifications
along closed boundaries.  Selected three-point functions
appear in 
\maybeeqs{\ref{eq:f111}-\ref{eq:f122}},
and we state the field identifications in
\maybeeqs{\ref{eq:height.one.identification}-\ref{eq:height.three.identification}}.

Next, in section~\ref{sec:Dissipation.General},
we introduce the concept of a
dissipative defect site, 
and discuss its
effect on the lattice Green functions
for the open, closed, and bulk cases.
In section~\ref{sec:Dissipation.Closed}, we 
show that 
in the closed and bulk cases,
dissipative defects have no effects on 
{\it any} weakly allowed cluster variables. This
demonstrates that an analysis of weakly allowed cluster
variables, such as that in~\cite{Mahieu.Ruelle},
cannot provide a complete picture of the ASM. 
Our results imply, as a particular case, that
dissipative defects in the closed and bulk cases
have no effect on the unit height probability,
or on correlations of unit heights.
They do, however,
have an effect on the higher height variables; we
show this analytically for the closed
case, in \maybeeqs{\ref{eq:defect2}-\ref{eq:defect3}},
and have checked this numerically for the
bulk case.

In sections~\ref{sec:all.open.correlations.complications},
\ref{sec:all.open.correlations.answer}, and 
\ref{sec:all.open.correlations.dissipation},
we compute all $n$-point correlation functions, for
any number of height variables, and with
any number of dissipative defects, along 
open boundaries. We find that there, all four
height variables, and dissipative defects,
are all represented by the same
dimension two field, $\lcftbdy$.
In fact, all local arrow diagrams along open
boundaries are represented by $\lcftbdy$, up to
multiplicative prefactors.

A short summary of these results can be found
at~\cite{Mine.Short}. 


\section{METHODS FOR ANALYSING THE ASM}
\label{sec:Green.Formalism}

At its core, the ASM is a tractable model because the
sandpile model has an Abelian structure; the state of the
sandpile does not depend on the order in which grains are
added to the sites~\cite{Dhar.First}. As a result of this Abelian
structure, it can be shown that
the states of the sandpile
fall into two simple categories.
Some of the $4^N$ states of the sandpile
(where $N$ is the number of sites) are transient,
which means that they can occur early in the ASM's
evolution, but occur with zero probability after an
infinitely long time. The other states are recurrent, and 
all occur with equal probability
after long times.
So the probability for a property X to occur is nothing more than the
fraction of recurrent states having property X.

To analyze the sandpile, it is convenient to allow
more general toppling rules. We characterize the sandpile
by a toppling matrix, $\DelSubs$, where
$\vi$ and $\vj$ are any lattice sites.
$\vi$ topples 
if its height is ever greater than $\DelSubsii$,
at which point its height goes down by 
$\DelSubsii$, and the height of 
every other site $\vj$ goes up by
$-\DelSubs \geq 0$ ($\DelSubs=0$ if $\vi$ and $\vj$
are not neighbors). The original ASM,
described in the introduction, has 
$\DelSubs =4$ when $\vi=\vj$ 
(or $\DelSubs =3$ when
$\vi=\vj$ is along
a closed boundary), 
$\DelSubs =-1$
when $\vi$ and $\vj$ are nearest neighbors, and
$\DelSubs =0$ otherwise. 

Dhar was able to show that the
number of recurrent states, given
very general restrictions on $\bf \Delta$,
is equal to
$\det ({\bf \Delta} )$~\cite{Dhar.First}. 
However, $\det({\bf \Delta})$
is also known to be 
equal to the number of spanning trees that
can be drawn on the lattice~\cite{Dhar.CFT}.
In the spanning tree representation,
$\DelSubsii$ indicates the number of 
neighbors that the arrow from $\vi$ can point to,
$\DelSubs =-1$ if an arrow can point from
$\vi$ to $\vj$, and $\DelSubs =0$ otherwise. 

Certain height probabilities 
in the ASM can be equated with probabilities for
spanning trees to have particular arrow
configurations. Probabilities for {\it some}
arrow configurations can be computed simply
by modfiying the toppling matrix from $\bf\Delta$
to $\bf \Delta'$,
in a way that 
enforces that arrow configuration. Then, the number of
spanning trees with the configuration is
$\det({\bf\Delta'})$, and the probability of 
the configuration is 
$\det ({\bf\Delta'})/\det({\bf\Delta})$.
Defining ${\bf B}\equiv{\bf\Delta'}-{\bf\Delta}$, 
the probability becomes

\begin{equation}
\frac{\det ({\bf\Delta'})}{\det({\bf\Delta})}=
\det(\id+{\bf BG}).
\end{equation}

\noindent ${\bf G}\equiv {\bf\Delta}^{-1}$ is 
the well-known lattice Green function~\cite{Spitzer}
(see appendix~\ref{sec:Green.Functions}).
If $\bf\Delta'$ only differs from $\bf\Delta$ in a finite
number of entries, then 
$\bf B$ is finite-dimensional, and
the probability can be easily computed.

Majumdar and Dhar used this method to find
the probability for a site $\vi$ to have unit 
height~\cite{Dhar.UnitCorrelations}. To
do this, they 
defined a modified, or ``cut'' ASM, in which 
three of 
the four bonds connecting $\vi$ to nearest neighbors 
are removed. When a bond is removed, the maximum height
of sites on each end is decreased by one; so the
three sites adjacent to $\vi$ get maximum heights of 3,
and $\vi$ gets a maximum height of 4-3=1. It is not 
difficult to show that recurrent states $S$ (of the original ASM)
where $\vi$ has height one are in
one-to-one correspondence with the recurrent
states $S'$ of the
cut ASM. In this correspondence, we map from $S$ to $S'$
by lowering the heights of each of the three
sites cut off from $\vi$ by one.
Letting $\vj_1$, $\vj_2$, and $\vj_3$, be the three
the neighbors that $\vi$ has been cut off from,
we have

\begin{eqnarray}
\nonumber
& & \quad\  \begin{array}{cccc}
\ \vi & \ \vj_1 & \ \vj_2 & \ \vj_3
\end{array} \\
{\bf B}  & = &
\left (
\begin{array}{cccc}
-3 & 1 & 1 & 1 \\
1 & -1 & 0 & 0 \\
1 & 0 & -1 & 0 \\
1 & 0 & 0 & -1
\end{array} \right)
\begin{array}{c}
\vi \\ \vj_1 \\ \vj_2 \\ \vj_3
\end{array}
\label{eq:BForHeightOne}
\end{eqnarray}

\noindent Then the unit height probability is
$\det (\id+{\bf BG}) = 2(\pi-2)/\pi^3$.
This method was also used by Majumdar and Dhar to calculate
the two-point correlation of the unit height 
variable~\cite{Dhar.UnitCorrelations}.

Priezzhev extended the Majumdar-Dhar 
method to allow for the
calculation of diagrams with closed loops.
With the basic Majumdar-Dhar method, all
off-diagonal entries of the
toppling matrix are either 0 or -1.
Priezzhev proved that if in $\bf\Delta'$ we set $n$
off-diagonal entries of $\bf\Delta$ 
to $-\epsilon$, then

\begin{equation}
\lim_{\epsilon\to\infty}
{{\det({\bf\Delta'})}\over{\epsilon^n}}
\end{equation}

\noindent is equal to the number of arrow
configurations such that 
each of the $n$ 
corresponding arrows is in a closed loop of arrows, where
each closed loop contributes a factor of $-1$,
and there are no closed loops
other than those going through these
$n$ bonds. 

Such configurations are not spanning
trees; spanning trees cannot have any closed loops.
However, Priezzhev found that to calculate 
certain spanning tree probabilities,
he needed to calculate graphs that
had closed loops ($\theta$-graphs). We
find this method useful for the
calculation of certain closed boundary correlations.


\section{HEIGHT PROBABILITIES}
\label{sec:Height.Probabilities}

Priezzhev determined a relationship between higher
height probabilities and spanning tree states, which
we review here~\cite{Priezzhev}.

Central to our analysis is the concept of forbidden
subconfigurations (FSCs). A forbidden subconfiguration is
a subset $F$ of the lattice, such that for all
$\vi\in F$, $h_{\vi} \leq c_{\vi}(F)$, where $h_{\vi}$ is the
height of site $\vi$, and $c_{\vi}(F)$ is number of neighbors
that $\vi$ has in $F$. Majumdar and Dhar proved
that a state of the ASM is 
recurrent if and only if it has no 
FSC's~\cite{Dhar.First,Dhar.CFT}.

The probability for a site $\vi$ to have height two is more
complicated than the height one probability~\cite{Priezzhev}. 
In this case, 
changing the site height to one  could either
leave the
ASM in an allowed (recurrent) state, or
produce an FSC. The first case just gives
the height one probability, which has
already been calculated, so
we consider the second case. Changing the
height of $\vi$ from two to one can produce multiple
FSC's. Let $F$ be the maximal forbidden subconfiguration (MFSC)
produced by this change. (Because more than one FSC can
be produced, the word ``maximal'' is necessary for
complete precision, and for this mapping to work; 
Priezzhev simply referred to
``the'' FSC, but this 
does not introduce any errors in his
analysis~\cite{Priezzhev}.)
$F$ must contain $\vi$, and exactly one of the
neighbors of $\vi$, and be simply connected,
but can otherwise have arbitrary shape. The states $S$ of
the original ASM where changing the height of $\vi$ from
two to one produces $F$ as the MFSC are in one-to-one
correspondence with states $S'$ of a modified ASM.
In the modified ASM,
all the bonds bordering $F$ are removed,
except for one (arbitrarily chosen) bond of $\vi$.
In the correspondence, we map from $S$ to $S'$
by lowering heights of all sites that border $F$ by
the number of neighbors of $F$ that they have been
cut off from.
(In this mapping,
heights in $F$ are unaffected.)
With this mapping,
the state $S$ has no FSC's  in the original ASM
if and only if
the state $S'$ has no FSC's in the cut ASM.
(Priezzhev's explanation used a slightly different, but
equivalent, argument, based on the burning algorithm, a 
method for determining if a state is 
recurrent~\cite{Dhar.First,Priezzhev}).

The site $\vi$ is called a predecessor of
the site $\vj$ in the spanning tree
if the path from $\vi$ to the root goes through
$\vj$.
We define $\NNP_{\vi}$, 
as the number of nearest-neighbors 
of $\vi$ that are predecessors
of $\vi$. 
Then, the correspondence above shows that
the number of states of the modified ASM is equal to the
number of spanning trees of the modified lattice, 
which is in
turn equal to the number of spanning trees where
$F$ is the set of predecessors of $\vi$.
Summing over all possible sets $F$, we 
simply obtain the number of spanning trees where
$\NNP_{\vi}=1$.

Similarly, it can be shown that the number of 
ASM states allowed when $\vi$ has height $h$ (or
greater) but forbidden when $\vi$ has height
$h-1$ (or less), is equal to the number of spanning trees where
$\NNP_{\vi}=h-1$. Thus, the probability $P_{ASM}(h)$
for the site to have exactly height $h$ in
the ASM is

\begin{equation}
\label{eq:Nonlocal.To.Height.One}
P_{ASM}(h) = \sum_{u=1}^h {{P_{SpTr}(u-1)}\over{m_{\vi}+1-u}}\ ,
\end{equation}

\noindent where $P_{SpTr}(u-1)$ is the
probability that a randomly chosen spanning 
tree will have
$\NNP_{\vi}=u-1$,
and $m_i$ is the maximum
possible height of $\vi$.
($m_{\vi}=4$ in the bulk, and along open boundaries,
while $m_{\vi}=3$ along closed boundaries.)
For more details, see~\cite{Priezzhev}.

This gives an exact representation of ASM height
probabilities in terms of spanning tree probabilities. 
However, these spanning tree probabilities are not easy
to calculate.
Spanning tree probabilities that correspond to
local restrictions on the spanning tree can be calculated 
with the Majumdar-Dhar method. However, the statement
that $\NNP_{\vi}=u-1$ is a {\it nonlocal}
restriction on the spanning tree (for $u>1$). 
Priezzhev was able to calculate these nonlocal
probabilities, but his calculations were complicated, and
do not appear to be easily extensible to calculation of
bulk correlations.
However, this problem turns out to be more tractable along 
a boundary.


\section{BOUNDARY HEIGHT PROBABILITIES}

\label{sec:Ivashkevich.Review}
\begin{figure}[tb]
\epsfig{figure=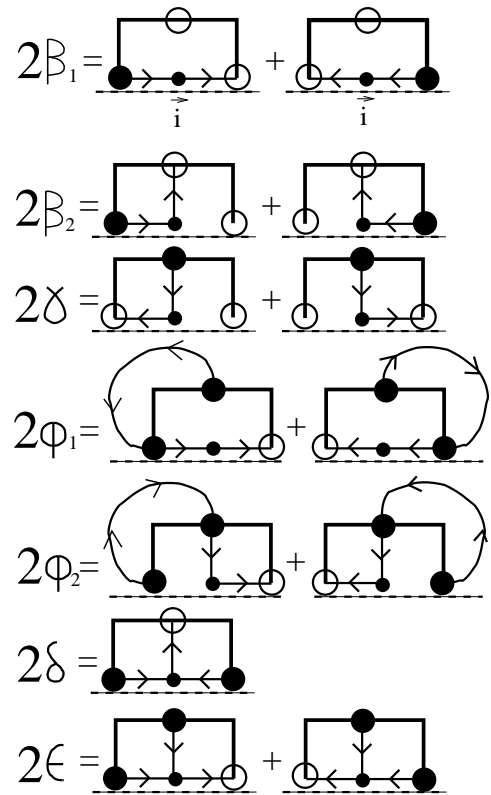,width=2.5in}
\caption{Nonlocal arrow diagrams along closed boundaries.}
\label{fig:NonlocalList}
\end{figure}

\begin{figure}[tb]
\epsfig{figure=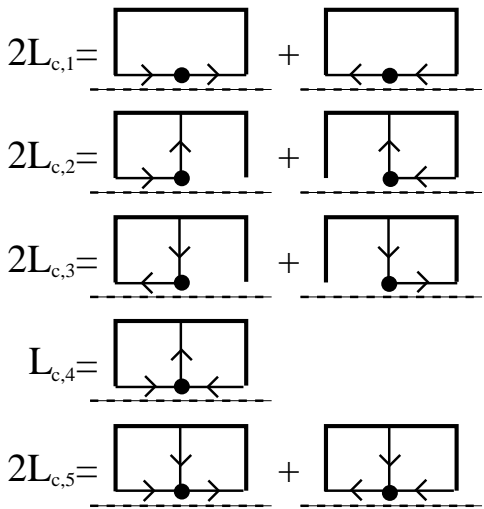,width=2.5in}
\caption{Local arrow diagrams along closed boundaries.}
\label{fig:LocalList}
\end{figure}

For sites at the boundary, the relationship between height 
probabilities and $\NNP$'s still holds, and the $\NNP$
condition is still nonlocal. Nevertheless, Ivashkevich was 
able to show, through an ingenious transformation, that 
the ASM height probabilities are much easier to calculate
along boundaries~\cite{Ivashkevich}. 

In figure~\ref{fig:NonlocalList} we list all possible 
nonlocal arrow configurations around a site $\vi$ of
a closed boundary. In each picture, 
the dashed line is the boundary,
and the central site
is $\vi$. Large, solid, circles
are predecessors of $\vi$, while
large, open, circles are not.
We see explicitly
that the predecessor relationships are nonlocal.
$\beta_1$ and $\phi_1$ differ only in whether the site
above $\vi$ leads to $\vi$ by a chain of arrows---since the
chain of arrows can go through sites distant from $\vi$,
this is a 
nonlocal distinction.
If we can figure out the
probabilities of all these diagrams, we can figure out 
the NNP probabilities (and thus the height probabilities).
For example, the probability 
for $\NNP_{\vi}=1$
is simply $2\beta_1+2\beta_2+2\gamma$, since
these diagrams catalogue all the ways that $\vi$
can have exactly one NNP. 

These nonlocal diagrams are difficult to calculate.
On the other hand, local restrictions
are easily calculated with the Majumdar-Dhar method.
All local arrow diagrams along closed boundaries 
are shown in figure~\ref{fig:LocalList}.
Note that these diagrams do not have solid or open circles,
because predecessor relationships are not 
specified in local diagrams.
Ivashkevich pointed
out that the local arrow diagrams could be written as
linear combinations of nonlocal arrow diagrams.
For example, looking at figures~\ref{fig:NonlocalList}
and~\ref{fig:LocalList}, we see that
$L_{c,1}=\phi_1+\beta_1$.
At first sight, there are
more nonlocal arrow diagrams than local arrow diagrams, 
so such linear relationships would not appear to let us solve for
the nonlocal arrow diagrams. However, Ivashkevich also pointed
out that certain nonlocal arrow diagrams are equal in
probability---for example, $\phi_1$ and $\phi_2$
are equal in probability, because we 
can make a one-to-one mapping from $\phi_1$ to $\phi_2$
by reversing all 
arrows in the long path of
$\phi_1$, and 
then switching the incoming arrow to $\vi$.
Similarly, $\beta_1=\beta_2$. Then, we
have as many nonlocal diagrams as local diagrams, and
can solve for the nonlocal arrow diagrams. 
(In fact, along open boundaries,
the number of local diagrams is one greater than the number
of nonlocal arrow diagrams,
so that the system is
overconstrained,
providing a check on the calculations.)
Ivashkevich used this to calculate all height
probabilities along open and closed boundaries.
See~\cite{Ivashkevich} for the full list of 
linear relationships between local and nonlocal 
diagrams.


\section{BOUNDARY TWO-POINT CORRELATIONS AND
ANOMALOUS GRAPHS}
\label{sec:Closed.Loop.Calculation}

The calculation of boundary correlations is much more 
difficult. We show in appendix~\ref{sec:Complication1}
that Ivashkevich's calculation of the two-point functions
was incorrect, and ignored complications that arise in the
relationship between ASM height correlations and spanning
tree correlations (although his end result turns
out to be correct).
In appendix~\ref{sec:Complication2}, we discuss further
complications that arise in transforming from nonlocal
spanning tree correlations to local spanning tree
correlations. We summarize the results here, and analyze
the resulting ``anomalous graphs''.

The first complication arises in the correspondence between
ASM height probabilities and spanning tree probabilities. It
would be natural to think that, analogously 
to \maybeeq{\ref{eq:Nonlocal.To.Height.One}},
the ASM probability, $P_{ASM}(h_{\vi},h_{\vj})$,
for the sites $\vi$ and $\vj$ to have exactly heights
$h_{\vi}$ and $h_{\vj}$ should be given by

\begin{equation}
\label{eq:Nonlocal.To.Height.Two}
P_{ASM}(h_{\vi},h_{\vj}) 
\begin{array}{c}
? \\ = \\ \\
\end{array}
\sum_{u=1}^{h_{\vi}}
\sum_{v=1}^{h_{\vj}}
{{P_{SpTr}(u-1,v-1)} \over {(m_{\vi}+1-u)(m_{\vj}+1-v)}} 
\ ,
\end{equation}

\noindent where $P_{SpTr}(u-1,v-1)$
is the probability that in a spanning tree,
$\NNP_{\vi}=u-1$ and $\NNP_{\vj}=v-1$.
However, this turns out to not be quite the
case. \Maybeeq{\ref{eq:Nonlocal.To.Height.Two}} 
is a natural guess, which we call a ``naive'' approach,
but as shown
in appendix~\ref{sec:Complication1}, the left and
right sides of \maybeeq{\ref{eq:Nonlocal.To.Height.Two}}
differ by a subset of spanning trees that we call anomalous
graphs of the first kind.
(These graphs are not anomalous in any physical
sense; we simply mean that they differ from what we would
get, using a certain naive starting point.)

\begin{figure}[tb]
\epsfig{figure=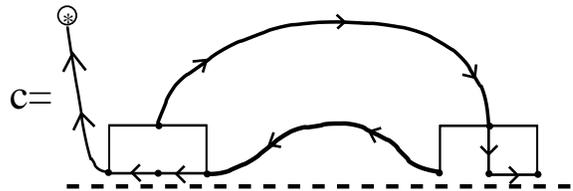,width=3.0in}
\caption{Anomalous graph of the first kind arising in the 
calculation of the two-point function.}
\label{fig:AnomalousGraphC}
\end{figure}

\begin{figure}[tb]
\epsfig{figure=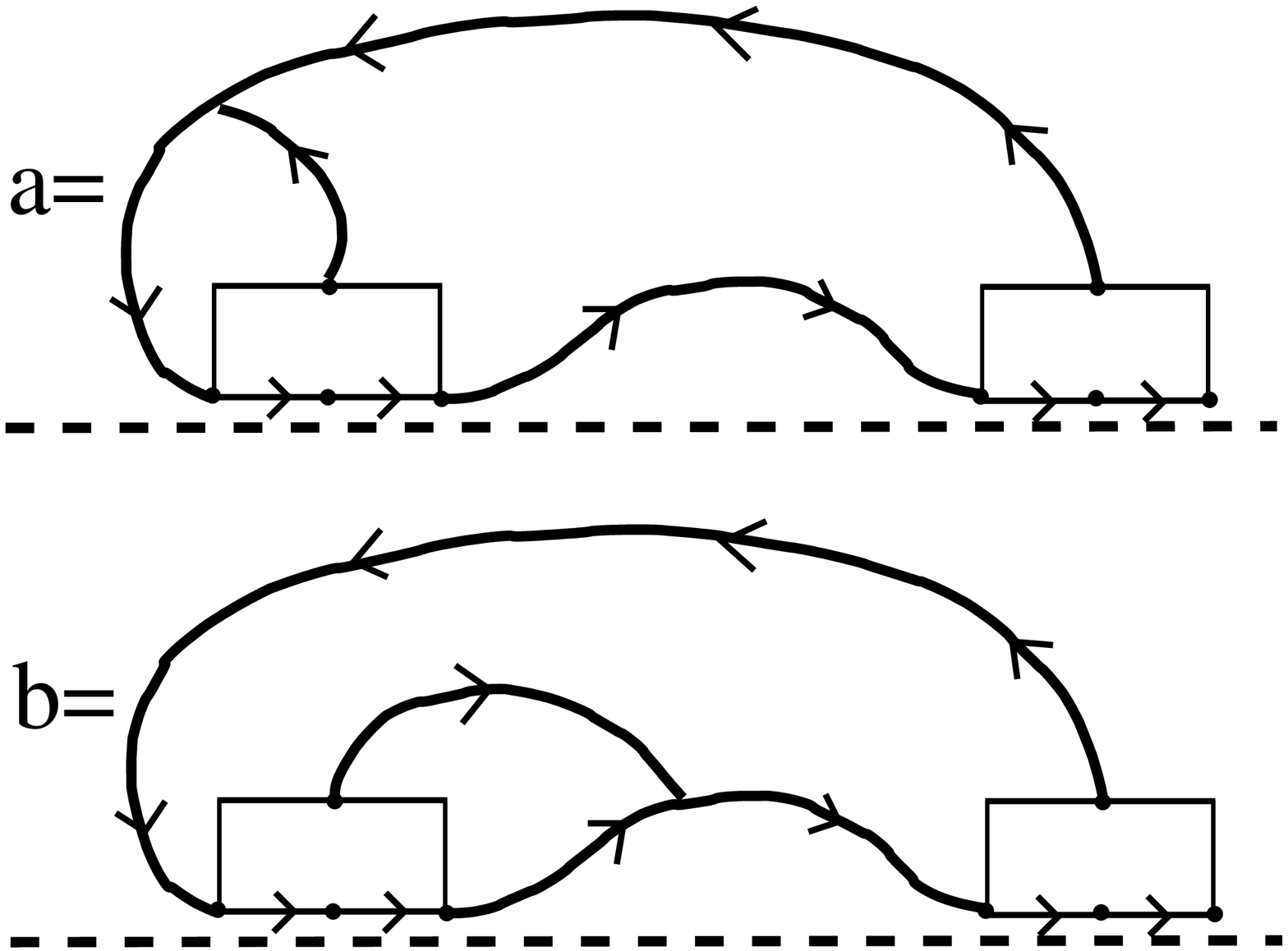,width=3.0in}
\caption{Anomalous graphs of the second kind arising in the 
calculation of the two-point functions.}
\label{fig:AnomalousGraphsAB}
\end{figure}

The set of anomalous graphs of the first kind,
for the closed case, is shown in
figure~\ref{fig:AnomalousGraphC}.
(We represent the root with a star.)
In the graph,
$\NNP_{\vi}=\NNP_{\vj}=1$, so this graph
appears in the right-hand side of 
\maybeeq{\ref{eq:Nonlocal.To.Height.Two}}
for $h_{\vi}=h_{\vj}=2$.
However, we show in appendix~\ref{sec:Complication1}
that this graph does not
contribute to the 2-2 (height two-height two) correlation,
but instead contributes to the 2-3 and 3-2 correlations, and
gets subtracted from the 3-3 correlation.

Second, leaving aside for now
the anomalous graphs of the first kind,
we need to
calculate correlations of nonlocal arrow diagrams. It would
be convenient if we could use the linear
relationships relating nonlocal arrow diagrams to local
arrow diagrams found for one-point functions
(section~\ref{sec:Ivashkevich.Review}),
and use them {\it independently} at
$\vi$ and at $\vj$ for two-point functions.
We again call this approach ``naive,''
and again, this approach does not quite work.
The problem arises because for one-point
functions, we treated $\phi_1$ and $\phi_2$
as equivalent, based on a one-to-one correspondence
in which a long path was reversed. In a correlation function 
of nonlocal arrow diagrams, the long path from a $\phi_1$
at $\vi$ may go
through arrow constraints near $\vj$,
which are not free to be reversed.
We discuss this problem in detail in 
appendix~\ref{sec:Complication2}.
Consideration of this problem 
shows that, relative to the naive approach,
our results are changed by graphs
$a$ and $b$, shown in 
figure~\ref{fig:AnomalousGraphsAB}.
We call these anomalous graphs of the second kind.

\begin{figure}[tb]
\epsfig{figure=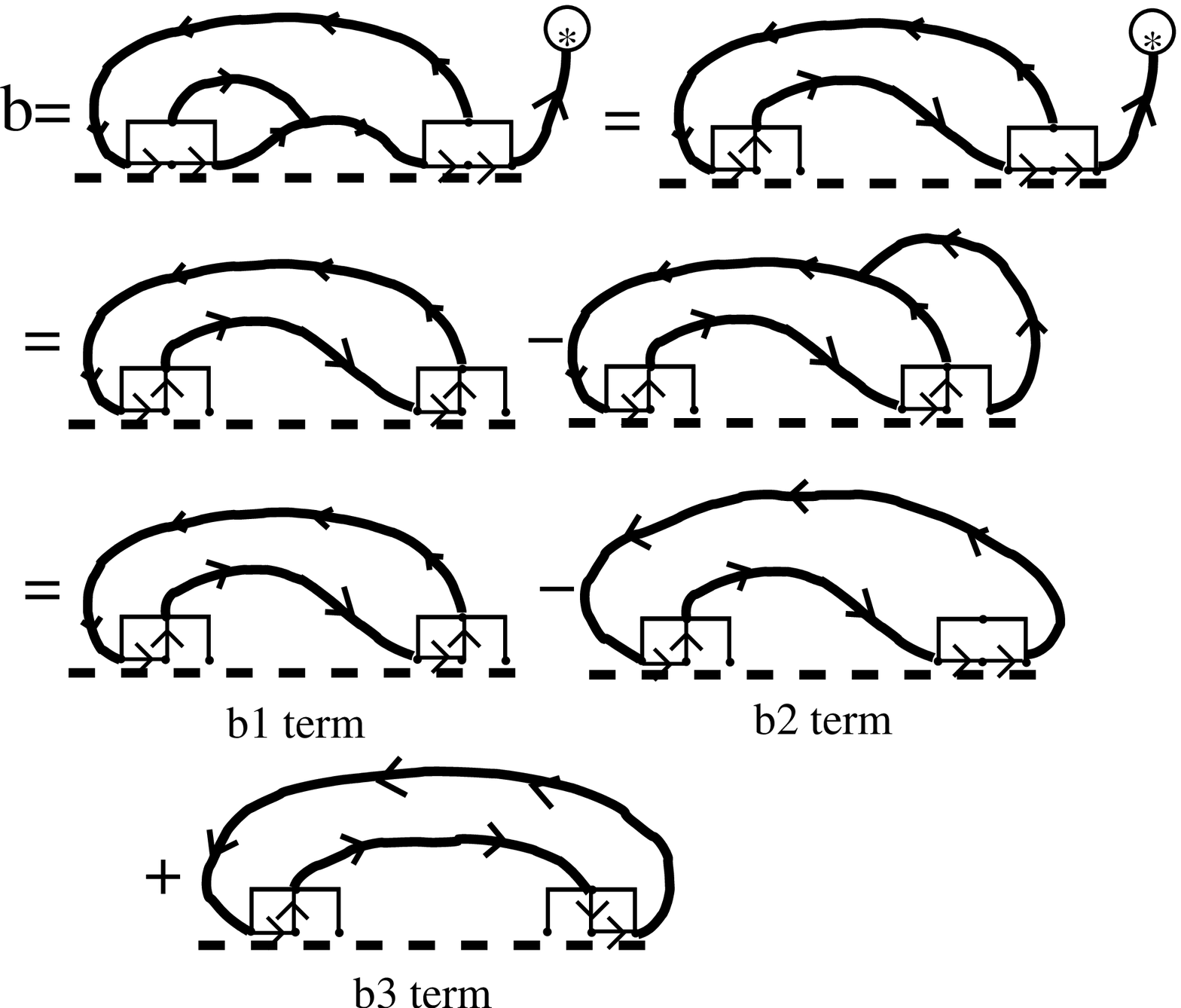,width=3.4in}
\caption{Anomalous graph $b$ as a linear combination of
closed loop diagrams.}
\label{fig:b.Closed.Loops}
\end{figure}

The anomalous graphs $a$, $b$, and $c$ can be
calculated with
the extension of the method of Priezzhev, discussed in
section~\ref{sec:Green.Formalism}~\cite{Priezzhev}.
We discuss only the calculation of the 
$b$ term; the analysis
of the other terms is similar.

$b$ represents a subset of spanning trees,
and thus
cannot have any closed loops. However,
it comes
``very close'' to having a closed loop that
includes the distant sites $\vi$ and $\vj$, and we 
see in figure~\ref{fig:b.Closed.Loops} that $b$ can be written
as a sum of closed loop diagrams. 

Priezzhev's method
allows us to calculate the closed loop diagrams.
We represent an arrow whose weight in
$\bf\Delta$ is set to $-\epsilon$ ($\epsilon\to\infty$)
with a wavy bond
line. As discussed in section~\ref{sec:Green.Formalism},
these bonds must be part of a closed loop,
and we get a factor of $-1$ for every closed loop.
This gives the relations in
figure~\ref{fig:Closed.Loops.General}.
Taking the difference of the two graphs
in figure~\ref{fig:Closed.Loops.General} 
then gives the value of a closed
loop diagrams that goes through both $\vi$ and $\vj$.
Using this method, we find the number of diagrams
$b1$, $b2$, and $b3$
(labeled in figure~\ref{fig:b.Closed.Loops}), 
as ratios of $N$, the total
number of spanning trees:

\begin{widetext}

\begin{eqnarray}
{N_{b1}\over N} & = & 
{{(3\pi-8)^2(\pi g(x)-1)}\over{4\pi^4x^2}} + 
{{(-128+48\pi+\pi^2)+(256-192\pi+30\pi^2)\pi g(x)}
\over{16\pi x^4}}
+\mathcal{O}({1\over x^6}) 
\label{eq:b1} \\
{N_{b2}\over N} & = & 
{{(3\pi-8)^2(\pi g(x)-1)}\over{2\pi^4 x^2}} +
{{(3\pi-8)(2\pi g(x) -1)}\over{4\pi^3 x^3}}+ 
{{(3\pi-8)((4-\pi)+(3\pi-8)\pi g(x))}
\over{2\pi^4x^4}}+
\mathcal{O}({1\over x^5}) 
\label{eq:b2} \\
\nonumber
{N_{b3}\over N} & = &
{{(3\pi-8)^2(\pi g(x)-1)}\over{4\pi^4 x^2}} +
{{(3\pi-8)(2\pi g(x) -1)}\over{4\pi^3 x^3}}+ 
{{(-128+48\pi-\pi^2)+(256-192\pi+42\pi^2)\pi g(x)}
\over{16\pi x^4}}+
\mathcal{O}({1\over x^5}) \\
\label{eq:b3}
\end{eqnarray}

\begin{figure}[tb]
\epsfig{figure=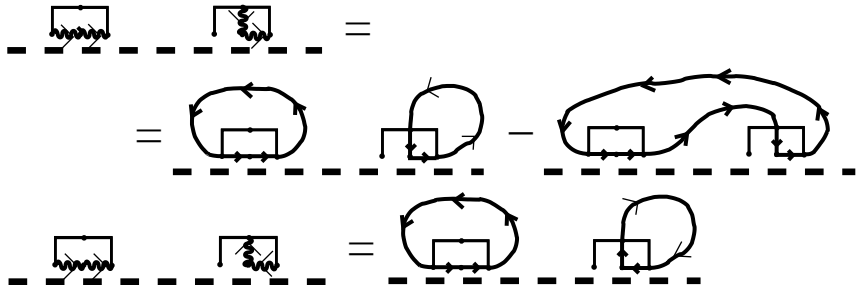,width=4.5in}
\caption{Use of -$\epsilon$ weight bonds to evaluate
closed loop diagrams.}
\label{fig:Closed.Loops.General}
\end{figure}

\end{widetext}

\noindent  $x$ is the separation between $\vi$ and $\vj$
along the defect. 
$g(x)$ is the Green function between $\vi$ and $\vj$,
and diverges
as $\ln (L)$, where $L$ is the system size
(it also diverges as $\ln (x)$).
The restriction that spanning trees should have no
closed loops
greatly limits the number of possible spanning trees, when
the outlets to the root (open boundaries)
are very far away. So diagrams such as
$b1$, $b2$, and $b3$, that allow a
closed loop, are much more numerous than 
diagrams of spanning trees.

However, to find $b$, we
take the linear combination,
$(N_{b1}-N_{b2}+N_{b3})/N$, and the $\ln(L)$
divergences cancel (this provides a check on our
calculations).
$a$ and $c$ can be
found similarly. We find

\begin{eqnarray}
\label{eq:a.value}
a & = & {{3\pi - 8} \over {2\pi^3 x^4}} + 
\mathcal{O} ({1 \over x^5}) \\
\label{eq:b.value}
b & = & {{3\pi - 8} \over {2\pi^3 x^4}} + 
\mathcal{O} ({1 \over x^5})  \\
c & = & \mathcal{O} ({1 \over x^6})
\end{eqnarray}

$a$ and $b$ are both of order $1/x^4$.
The two-point correlation functions turn out to decay as
$1/x^4$, so the anomalous graphs could, in principle,
affect the universal parts of the correlation
functions. However, 
the anomalous graphs of the second kind
come in the combination $(a-b)$ 
(see \maybeeqs{\ref{eq:ab.contribution.1}-\ref{eq:ab.contribution.3}}).
So their total contributions to the two-point
correlations are $\mathcal{O}(1/x^5)$, and 
can be dropped.

The end result is rather surprising. A ``naive''
approach might simply apply the relationship between
ASM states and NNP conditions found for the 
one-point functions, independently at $\vi$ and $\vj$
(i.e. extend \maybeeq{\ref{eq:Nonlocal.To.Height.One}} to
\maybeeq{\ref{eq:Nonlocal.To.Height.Two}}), and then apply
the relationships between nonlocal arrow diagrams and local
arrow diagrams found for the one-point functions,
independently at $\vi$ and $\vj$.
Neither of these steps is correct, and a correct
analysis produces correction terms (the anomalous
graphs) to this naive approach. But, somehow, the 
anomalous graphs, while nonzero, produce no correction to
the leading-order, universal results at any stage of the
computation; the naive approach gives the answers.
In fact, we find in the following
sections that the naive approach again gives correct results
for all three-point
closed boundary correlations that we have calculated, and
for all open boundary correlations. This leads
us to conjecture that the naive approach always produces
correct universal results, for all correlations. If this
conjecture were proven true, it would greatly simplify
further calculations---for example, the anomalous 
diagrams have
prevented us from calculating the 2-2-2 correlation
along closed boundaries.


\section{TWO- AND THREE-POINT CLOSED BOUNDARY CORRELATION FUNCTIONS}
\label{sec:Three.Point.Closed}

We define, for all 
correlation functions along closed boundaries,

\begin{eqnarray}
\nonumber
f_\cc (a_1,a_2,\dots,a_n) & = &\\
& & 
\hspace{-1.0 in} 
\langle (\delta_{h_{x_1},a_1}-p_{a_1,\cc})
\dots
(\delta_{h_{x_n},a_n}-p_{a_n,\cc}) \rangle_\cc \ ,
\label{eq:closed.corr.def}
\end{eqnarray}

\noindent In this correlation function, the
height $h_{x_u}$ at the boundary site $x_u$ is 
required to be $a_u$. 
We have subtracted off the 
constant boundary probabilities, 
$p_{a_u,\cc}$,
which were found in~\cite{Ivashkevich},
as described in section~\ref{sec:Ivashkevich.Review}.
The subscript ``$\cc$'' stands for ``closed.''
As already noted,
despite errors in the setup 
in~\cite{Ivashkevich}, the results
of~\cite{Ivashkevich} are nevertheless
correct, where it was found that

\begin{eqnarray}
\label{eq:twoPoint.11}
f_\cc (1,1) & = & \left(-{9 \over \pi^2} + {48 \over \pi^3} 
- {64 \over \pi^4} \right) {1 \over {(x_1-x_2)^4}} +\dots \\
\label{eq:twoPoint.12}
f_\cc (1,2) & = & \left({12 \over {\pi^2}} - {68 \over \pi^3} 
+ {96 \over \pi^4} \right) {1 \over {(x_1-x_2)^4}} +\dots \\
\nonumber
f_\cc (2,2) & = & \left(-{61 \over {4 \pi^2}} + {96 \over \pi^3} 
- {144 \over \pi^4} \right) {1 \over {(x_1-x_2)^4}} +\dots \\
\label{eq:twoPoint.22}
\end{eqnarray}

\noindent The correlation functions involving the height
three variables were also calculated, but we do not write
them here, as they are determined
by the requirement that all
height probabilities must sum to one at every
site. (There is a misprint in the result for
$f_\cc (3,3)$ in~\cite{Ivashkevich}.)

Ivashkevich argued that the fact that all 
two-point correlations
decay as $1/(x_1-x_2)^4$ indicates that all
three height variables are represented by the same
field operator. However, if all height variables were
represented by the same operator, we would expect
the two-point functions to factorize, as

\begin{equation}
\label{eq:factorize}
f_\cc (u,v) = - {{K_u K_v}\over{(x_1-x_2)^4}} \quad ; \quad
u,v\in\{1,2,3\} \ ,
\end{equation}

\noindent for some constants $K_u$.
However, the results in 
\maybeeqs{\ref{eq:twoPoint.11}-\ref{eq:twoPoint.22}}
do not factorize in this manner. 
Dhar argued that we should expect this factorization 
for bulk correlations, based on the ``clustering
properties of correlation functions,'' but we see that this
factorization already fails along closed 
boundaries~\cite{Dhar.AllSame}.
(We will see later that
the open boundary correlations
do, however, factorize in this manner, for
all $n$-point correlations.)

To clarify the field identifications, we have calculated
all three-point functions along closed boundaries, where
at least one of the heights is the unit height. 
Some of our results are

\begin{eqnarray}
\nonumber
f_\cc (1,1,1) & = & {{2(3\pi-8)^3} \over
{\pi^6 (x_1-x_2)^2(x_1-x_3)^2(x_2-x_3)^2}}
+\dots \\
\label{eq:f111}
\\
\nonumber
f_\cc (1,1,2) & = & - {{8(\pi-3)(3\pi-8)^2} \over
{\pi^6 (x_1-x_2)^2(x_1-x_3)^2(x_2-x_3)^2}} \\
& & - {{(3\pi-8)^2} \over {\pi^5
(x_1-x_3)^3(x_2-x_3)^3}}+\dots \\
\nonumber
\label{eq:f122}
f_\cc (1,2,2) & = & - {{4(3\pi-8)(-5\pi^2+39\pi-72)} \over
{\pi^6 (x_1-x_2)^2(x_1-x_3)^2(x_2-x_3)^2}} \\
& & + {{(3\pi-8)(24-7\pi)} \over {2\pi^5 (x_1-x_2)^3(x_1-x_3)^3}}
+\dots
\end{eqnarray}

Other three-point correlation functions, calculated with the
same methods, are listed in
appendix~\ref{sec:Additional3pts}. They are all 
consistent with the requirement that the three height
probabilities must sum to one at any site, and permutation
symmetry, thus providing a check on our calculations.

We again get a number of anomalous graphs (relative to
a naive approach), and as stated in the previous section,
again find that all anomalous graphs cancel in the
universal,
leading-order terms of the correlation function.

These correlation functions are
consistent with identifying the height variables
with the following
field operators in the $c=-2$ CFT:

\begin{eqnarray}
\label{eq:height.one.identification}
{\rm  Height\ one} & : &
-{{2(3\pi-8)}\over {\pi^2}}
\left( \partial\theta\partial\bar\theta \right) \\
\label{eq:height.two.identification}
{\rm Height\ two} & : &
{{6(\pi-4)}\over {\pi^2}}
\left( \partial\theta\partial\bar\theta \right)
+{1\over{2\pi}} \theta\partial^2\bar\theta \\
\label{eq:height.three.identification}
{\rm Height\ three} & : &
{8\over {\pi^2}}
\left( \partial\theta\partial\bar\theta \right)
-{1\over{2\pi}} \theta\partial^2\bar\theta
\end{eqnarray}

\noindent The representation
of the $c=-2$ CFT used here is
described briefly in appendix~\ref{sec:c-2LCFT}.
Note that the boundary correlations in
\maybeeqs{\ref{eq:f111}-\ref{eq:f122}}
are the same as the {\it bulk} correlations
of \maybeeqs{\ref{eq:height.one.identification}-\ref{eq:height.three.identification}},
and that while the $c=-2$ CFT contains
holomorphic and antiholomorphic fields
(the $\partial$ and $\bar\partial$ of 
\maybeeq{\ref{eq:LCFT.action}}), the fields in
\maybeeqs{\ref{eq:height.one.identification}-\ref{eq:height.three.identification}}
contain only holomorphic fields.
This is consistent with boundary 
CFT.
While fields in the bulk generally 
have holomorphic and antiholomorphic
parts, near a boundary the antiholomorphic
pieces behave, in all correlation functions, like
holomorphic pieces at mirror positions
across the boundary~\cite{Cardy.mirror}.

It is also consistent to make the substitution
$\theta\to\bar\theta$, $\bar\theta\to-\theta$ in
these field identifications,
as the $c=-2$ LCFT is symmetric under this 
transformation (see \maybeeq{\ref{eq:LCFT.action}}).

The fact that the field identifications
for the height variables differ
along a closed boundary proves that they must also differ
in the bulk. This is because in a CFT, boundary
operators are derived from operator
product expansions (OPE's) of 
bulk operators~\cite{Cardy.mirror}. 
Furthermore, in appendix~\ref{sec:Simple.Proof}
we present a simple argument, based on general 
CFT principles, and not on any detailed calculations, that 
the height variables must have different field
identifications in the bulk.

We have not been able to calculate three-point 
correlation functions that have no unit height
variables. The basic problem is with the anomalous
diagrams that arise when we 
convert from nonlocal arrow diagrams
to local arrow diagrams (as in
appendix~\ref{sec:Complication2}).
The trick shown in
figure~\ref{fig:Closed.Loops.General},
for evaluating
the resultant
closed loop diagrams,
does not work for these three-point functions.
We note that if we use the conjecture
proposed in section~\ref{sec:Closed.Loop.Calculation}
(i.e., ignore the anomalous graphs), we obtain

\begin{equation}
\label{eq:f222}
f_\cc (2,2,2) = -  {{(24-5\pi)(-576+384\pi-61\pi^2)} \over
{4\pi^6(x_1-x_2)^2(x_1-x_3)^2(x_2-x_3)^2}} +\dots,
\end{equation}

\noindent (and other three-point functions consistent
with the requirement that all three height probabilities
must sum to one at any site). This correlation
function is consistent with the 
field identification in 
\maybeeq{\ref{eq:height.two.identification}},
providing support for our conjecture.


\section{DISSIPATIVE DEFECT SITES, GENERAL}
\label{sec:Dissipation.General}

We now consider the effects of dissipative defects on the
ASM. Generally, at sites in the bulk, or along
closed boundaries,
the number of grains is conserved at each toppling.
Usually, it is only at open boundaries that
the number of grains is not conserved;
there, of the four grains toppled, three grains are sent
to neighbors, while the remaining grain goes
off the edge of the sandpile, to the root. 

Some dissipation (i.e. sites where
topplings remove grains
from the sandpile) is necessary for the sandpile
model to be well defined, since otherwise we would
end up with states where the topplings never terminated.
Nevertheless, dissipation often plays a 
minor role in analysis of the sandpile,
because properties are often studied
in the bulk of the ASM, with the 
dissipative sites along the open boundaries infinitely far
away.

Some previous studies have investigated the effect of adding
dissipation throughout the bulk of the ASM.
Instead of having bulk sites topple when their height
is greater than 4, they topple when their height is
greater than $4+k$ ($k>0$). Then, one grain is sent to each
of the four neighbors, and $k$ grains are lost 
to the root. 
It has been shown, both numerically and
analytically, that if this is done at all sites,
the ASM is taken off the critical 
point, and the power law correlations
are destroyed~\cite{DissipBreaksSOC,dissip.1,dissip.2}.
This happens even when $k$ is 
infinitesimal. (Although this modification to the
ASM has its most obvious interpretation for 
integer $k$, the theory can be given a sensible
interpretation for any rational value of $k$.
See~\cite{DissipBreaksSOC} for details.)
More recently, Mahieu and Ruelle have demonstrated the
precise manner in which dissipation throughout the
bulk takes the ASM off the critical point. They
found that the dissipation has exactly the same effect on 
correlation functions of weakly allowed cluster
variables, as adding 
the integral of the dimension 0 variable,
$\theta\bar\theta$, to the $c=-2$ CFT~\cite{Mahieu.Ruelle}.
Adding dissipation along a line has been shown
to split the ASM into two separate half-planes,
each with open boundary 
conditions~\cite{Mine.BC.Univ}.

Here, we consider the effect of adding dissipation
at only a single defect site. 
Then, the methods of Majumdar and Dhar still work, but
we need to use a modified
lattice Green function.  
If $k$ grains of sand are dissipated at the lattice
position $\vec{d}$, then we call $k$ the ``strength''
of the defect. The toppling matrix is then
changed from the defect-free toppling matrix,
${\bf\Delta}_0$, to

\begin{equation}
\label{eq:defect.general}
\Delta_{\vi,\vj} =
\Delta_{0;\vi,\vj}+
k \delta_{\vi,\vec{d}} \; \delta_{\vj,\vec{d}}
\end{equation}

\noindent The Green function is simply the inverse of the
toppling matrix, and is changed from the defect-free
Green function, $G_0$
(described in appendix~\ref{sec:Green.Functions}), to

\begin{equation}
\label{eq:defect.Green}
G (\vi,\vj) = 
G_0 (\vi,\vj) -
{k \over {1+k G_0 (\vec{d},\vec{d})}}
G_0 (\vi,\vec{d}) G_0 (\vec{d},\vj)
\end{equation}

\noindent This holds for any value of $k$, and regardless of
the location of the defect. Nevertheless, the defect
behaves very differently in the open case, and 
in the closed/bulk cases. This is because the 
Green function between nearby lattice 
sites is $\mathcal{O} (1)$ near
an open boundary, but 
$\mathcal{O} (\ln L)$ 
near a closed boundary, or in the
bulk~\cite{Spitzer,Bdy.Falloff}.
$L$ is the system size, or more generally,
is of the same order-of-magnitude as 
the distance to the nearest open 
boundary. 
This divergence in the Green function as $L\to\infty$ for
the closed and bulk cases is usually not an issue,
since in most cases, we are concerned with
differences in Green functions. 
However, here the divergence of all the
${\bf G}_0$ terms makes 
\maybeeq{\ref{eq:defect.Green}} unwieldy, although 
technically correct. (\maybeeq{\ref{eq:defect.Green}}
can be used in the open case without modification.)
We work in a limit where the distances between
$\vi$, $\vj$, and $\vec{d}$,
while possibly large, are all
much less than $L$.
In this limit, dropping terms of
order $1/(\ln L)$, \maybeeq{\ref{eq:defect.Green}}
becomes

\begin{equation}
\label{eq:defect.bulk.closed}
G(\vi,\vj)=
G_0 (\vi,\vj) - G_0 (\vi,\vec{d})
-G_0 (\vec{d},\vj) + G_0 (\vec{d},\vec{d})
\end{equation}

\noindent Note that
\maybeeq{\ref{eq:defect.bulk.closed}}
is independent of $k$. This makes sense, since
in the bulk, or along a closed boundary,
spanning trees have to travel far to reach the root.
But with the defect given by
\maybeeq{\ref{eq:defect.Green}}, $k$ bonds are added from the
defect $\vec{d}$ to the root.
Adding a dissipative defect provides such an ``easy''
way to reach the root, that with high probability
(probability one as $L\to\infty$), all nearby points will
be predecessors of the dissipative defect, regardless
of the value of $k$. The set of spanning trees will thus
be the same, in the $L\to\infty$ limit, for any $k$.
Note also that the Green function in
\maybeeq{\ref{eq:defect.bulk.closed}} 
no longer diverges as $L\to\infty$,
which is appropriate, as we are no longer 
$\mathcal{O}(L)$ from any dissipative sites.


\section{DISSIPATIVE DEFECT SITES, CLOSED AND BULK CASES}
\label{sec:Dissipation.Closed}

Surprisingly, it turns out that a dissipative defect,
either in the bulk, or on or near a closed
boundary, has no effect on any 
weakly allowed cluster variables in the ASM.
Weakly allowed cluster variables are 
height configurations that result in a 
subconfiguration that contains an FSC if any
height in the configuration is reduced
by one~\cite{Majumdar.Dhar.II}. Examples of 
weakly allowed cluster variables are a single height
one variable, or a height one adjacent to a
height two. 
Such variables can be calculated with the Majumdar-Dhar
method by the removal of a set of bonds in the
ASM/spanning tree.
We note that correlations of weakly allowed
cluster variables (such as all correlations of 
the unit height variable) are also weakly allowed
cluster variables.

Probabilities of weakly allowed cluster
variables
can be calculated as $\det (\id+{\bf BG})$,
as in section~\ref{sec:Green.Formalism}.
To analyze the effects of the defect, we want to
consider the effect of
modifying the Green function 
from the defect-free Green function ${\bf G}_0$,
to the Green function in
\maybeeq{\ref{eq:defect.bulk.closed}},
for a fixed matrix $\bf B$ (i.e. for a specific
weakly allowed cluster variable).

In general, for local arrow restrictions, each row
of $\bf B$ must sum to zero, because if
the restrictions on the spanning trees prevent an
arrow from $\vi$ to $\vj$, then
$B_{\vi ,\vi}$ goes down by 1, while $B_{\vi ,\vj}$ goes up 
by 1. 
(For example, for the height one variable, the
matrix $\bf{B}$ in \maybeeq{\ref{eq:BForHeightOne}}
arises from the restriction that no arrows can 
point from $\vi$ to $\vj_1$, $\vj_2$, or $\vj_3$,
nor from $\vj_1$, $\vj_2$, or $\vj_3$ to $\vi$.)
For the weakly allowed cluster variables,
$\bf B$ is symmetric, since
if the arrow from $\vi$ to
$\vj$ is forbidden, then so is the
arrow from $\vj$ to $\vi$.
So every column of $\bf B$ also sums to zero.

Since every row of $\bf B$ sums to zero, the parts of 
$\bf G$ that are independent of the row index of 
$\bf G$ make
no contribution to $\bf BG$, and thus no contribution
to the probability $\det (\id+{\bf BG})$.
And $\det (\id+{\bf BG})=\det (\id+{\bf GB})$,
so since every column of $\bf B$ sums to zero, the
parts of $\bf G$ that are independent of the column
index also make no contribution to the probability.
The last three terms of 
\maybeeq{\ref{eq:defect.bulk.closed}}
all depend either only on the row index, or only on the
column index. So a dissipative defect has no
effect on any weakly allowed cluster probabilities
(either on one-point probabilities or on correlations).

As a special case, this
means that the unit height probability, and its
correlations, are unaffected by closed or bulk
dissipative defects. 
However, the higher height variables
are affected. Using the Green function in 
\maybeeq{\ref{eq:defect.bulk.closed}},
and the methods described in
section~\ref{sec:Ivashkevich.Review},
we find
that along a closed boundary, with a defect
at the origin, we have the following height
probabilites at $x_1$:

\begin{eqnarray}
\label{eq:defectnochange}
f_\cc (1) & = & 0 \\
\label{eq:defect2}
f_\cc (2) & = & - {1 \over {2\pi x_1^2}} + \dots \\
\label{eq:defect3}
f_\cc (3) & = & + {1 \over {2\pi x_1^2}} + \dots 
\end{eqnarray}

\noindent We have numerically confirmed these results.
These results provide further evidence that
the height two and three variables have different
field identifications along closed boundaries.

Since the height two and three variables have
dimension two,
this indicates that a 
dissipative defect along a closed boundary is
a dimension zero operator.
Consistent with this,
uniform dissipation in the bulk has been
identified with the integral of the dimension zero
operator $\theta\bar\theta$~\cite{Mahieu.Ruelle}.
However, the correlation of
$\theta\bar\theta$ with the height two and 
three operators in 
\maybeeqs{\ref{eq:height.two.identification}-\ref{eq:height.three.identification}}
does not produce the correlations in
\maybeeqs{\ref{eq:defectnochange}-\ref{eq:defect3}};
this situation requires further analysis.

In the bulk, we would also expect that the
higher
height probabilites would 
be affected by a defect site, and have confirmed this
with numerical simulations, although have not proven
this analytically.

The fact that weakly allowed cluster variables have no
correlations with bulk or closed defects provides 
compelling evidence that weakly allowed cluster
variables do not provide a complete picture of
the sandpile model. This has particular
bearing on the analysis of Mahieu and
Ruelle~\cite{Mahieu.Ruelle}. They studied specific
bulk correlations of the
simplest weakly allowed cluster variables,
and developed a 
complete field picture for these variables.
They found that (at the critical point)
these variables are all linear
combinations of three dimension two variables,
$\lcftbulk$, 
$\partial\theta\partial\bar\theta$, and
$\bar\partial\theta\bar\partial\bar\theta$,
strongly indicating that all
weakly allowed cluster variables 
are linear combinations of these
three fields. 
However, this analysis left the
status of the
height two variable unresolved.
Mahieu and Ruelle pointed out
that since the height two variable
appears in a number of the weakly allowed cluster
variables, it might be expected that the 
height two variable would also be a linear
combination of these three fields, or more
specifically, proportional to 
the sole rotationally invariant
field, $\lcftbulk$~\cite{Mahieu.Ruelle}. But they
also noted that such an identification
appeared inconsistent with the fusion rules of the $c=-2$
CFT, which would indicate
a different field identification.
The analysis here points 
strongly to the latter conclusion, although the
specific field identification in the bulk 
remains unresolved.


\section{ALL \protect\boldmath$n$-POINT CORRELATIONS ALONG
OPEN BOUNDARIES, PART I}
\label{sec:all.open.correlations.complications}

We have calculated all $n$-point correlations of all
four height variables, along open boundaries, in 
the presence of an arbitrary number of dissipative 
defects. We begin by discussing why this case is so
tractable (in contrast to the closed case, where
we have been unable to calculate the three-point
function of the height two variable).

The heights of the correlation function are
placed at $x_1, x_2, \cdots, x_n$,
and defining
$x_{ab}\equiv x_a-x_b \equiv c_{ab} x$, we work in  
the limit $x\to\infty$, where the
$c_{ab}$'s are kept constant.

As discussed
in section~\ref{sec:Closed.Loop.Calculation}, and 
appendices~\ref{sec:Complication1}
and~\ref{sec:Complication2}, 
a number of anomalous terms
arise in the computation of
correlation functions. 
While the discussion in these sections
focused on closed boundary correlations, similar anomalous
graphs arise in open boundary correlations.
However, it turns out that these
anomalous graphs produce no contributions to the universal
parts of any correlation functions,
greatly simplifying matters.
We prove this claim in this section, and in
the next section look at the actual calculation of
the correlation functions.

We start by focusing on the two-point correlations.
Note that the 
anomalous graphs found thus far, in
figure~\ref{fig:AnomalousGraphC}
and~\ref{fig:AnomalousGraphsAB}, 
all involve ``nearly-closed''
loops: the trees have paths that go from the
neighborhood of $\vi$ to the neighborhood of $\vj$, and
from the neighborhood of $\vj$ to the neighborhood 
of $\vi$.
The paths do not actually form closed loops,
since no closed loops are allowed in spanning
trees, but they do come very close (within one site).
The reasons for this are general, so similar structures
will arise in all anomalous graphs, for
all correlation functions.
For example, the 
anomalous graphs in figure~\ref{fig:AnomalousGraphsAB}
arose because a long, nearly-closed loop from 
one site
could not be
reversed in direction, 
if it passed through fixed arrows 
at the other site
(see figure~\ref{fig:Phi1.Phi2.Complications} of
appendix~\ref{sec:Complication2}).

In the open case, these anomalous graphs
between $\vi$ and $\vj$ always fall off faster than
$\mathcal{O}(1/x^4)$. 
This is in contrast to the closed boundary case, where such
diagrams diverge---see 
\maybeeqs{\ref{eq:b1}-\ref{eq:b3}}.
The difference results from the Green functions. While
the Green function diverges as $\ln (x)$ along
closed boundaries, it decays as $1/x^2$ along open
boundaries
(see appendix~\ref{sec:Green.Functions}).
Using Priezzhev's method,
the matrix determinant for evaluating any closed
loop diagrams
necessarily 
involves two Green functions, one from $\vi$ to $\vj$, and
another from $\vj$ to $\vi$, giving an overall 
factor of $1/x^4$.
Furthermore, calculating the
diagrams requires two
matrix determinants, which come with leading terms
equal in magnitude, but opposite
in sign---see figure~\ref{fig:Closed.Loops.General}.
The $\mathcal{O}(1/x^4)$
parts of the closed loop diagrams thus cancel
along open boundaries. So
the anomalous graphs
for the two-point functions automatically
fall off faster than $\mathcal{O}(1/x^4)$,
and do not
need to be considered when calculating
leading-order, universal parts of correlation
functions.

By this logic, for any $n$-point open boundary
correlations, any
anomalous graphs must have
terms that decay as $1/(x_a-x_b)^p$, where
$p\geq 5$, for some $a,b\in 1,2,\cdots n$. 
Aside from the sites at $x_a$ and $x_b$, there are
$(n-2)$ other sites
that need to appear in the connected correlation function.
Each brings a new
Green function, of $\mathcal{O}(1/x^2)$, so
the overall
contribution of any anomalous graph must
decay at
least as fast as $\mathcal{O}(1/x^{5+2(n-2)})=
\mathcal{O}(1/x^{2n+1})$. But we will see in the next section
that all $n$-point correlations 
decay to leading order as $1/x^{2n}$. So the
anomalous graphs have
no effect on the universal parts of any $n$-point
correlation functions.
The conjecture at the end of 
section~\ref{sec:Closed.Loop.Calculation}
has thus been
proven for all open boundary correlations.


\section{ALL \protect\boldmath$n$-POINT CORRELATIONS ALONG
OPEN BOUNDARIES, PART II}
\label{sec:all.open.correlations.answer}

Since we can ignore the anomalous graphs
for open boundary correlation functions,
no error is introduced
by writing the height probability at each site
as a linear combination of local arrow diagrams,
independently
using at each site
the linear relationships derived 
for the one-point functions.
Defining the open boundary correlation $f_\op$
analogously to $f_\cc$ for the closed 
case (\maybeeq{\ref{eq:closed.corr.def}}), we
then have

\begin{eqnarray}
\nonumber
f_\op (a_1,a_2,\dots,a_n) & = & \\
\nonumber
& & \hspace{-1.0in} =
\sum_{u_1=1}^{N_{\loc}} 
\sum_{u_2=1}^{N_{\loc}} \dots
\sum_{u_n=1}^{N_{\loc}} 
D_{a_1u_1}D_{a_2u_2}\dots D_{a_nu_n} \\
& & \hspace{-1.0in}
\qquad \left<
L_{\op,u_1}(x_1) L_{\op,u_2}(x_2) \dots L_{\op,u_n}(x_n)
\right>
\label{eq:open.height.to.local}
\end{eqnarray}

\noindent Each $L_{\op,u_f}(x_f)$
represents a local arrow diagram at $x_f$,
analogous to the diagrams in 
figure~\ref{fig:LocalList},
but for the open case,
and $N_{\loc}$ is the total number of possible local arrow
diagrams at a single site
(see~\cite{Ivashkevich} for the list of diagrams).
$\bf D$ is a constant matrix expressing
height probabilities in terms of local arrow diagrams,
for one-point functions, and
was (implicitly)
found in~\cite{Ivashkevich}.
Each correlation of local arrow diagrams can now be
calculated with the Majumdar-Dhar method.

If a site $\vi$ has local arrow constraints $u$,
we express those constraints by a matrix ${\bf B}_u$,
and let ${\bf G}_{uu}$ be the
Green function matrix for the sites around $\vi$.
${\bf B}_u$ and ${\bf G}_{uu}$ are both associated only
with sites in the vicinity of $\vi$.
$p_u=\det (\id +{\bf B}_u {\bf G}_{uu})$
gives the one-point probability for the local arrow
diagram $L_{\op,u}$.
The two-point correlation of local
arrow diagrams $u_1$ and $u_2$ is given by
$\det (\id +{\bf BG})$, where
${\bf B}$ is block diagonal, with ${\bf B}_{u_1}$
and ${\bf B}_{u_2}$ along the block diagonal, and
${\bf G}$ is made of the four matrix blocks
${\bf G}_{u_1u_1}$, ${\bf G}_{u_1u_2}$,
${\bf G}_{u_2u_1}$, and ${\bf G}_{u_2u_2}$.
Mahieu and Ruelle found that the
leading order contribution to the 
{\it bulk} two-point probability is given 
by~\cite{Mahieu.Ruelle}

\begin{widetext}

\begin{equation}
\label{eq:MR.Trace.2}
\det (\id+{\bf BG}) = 
-p_{u_1}p_{u_2} {\rm Trace}
\left\{
{{\id}\over{\id+{\bf B}_{u_1}{\bf G}_{u_1u_1}}}
{\bf B}_{u_1} {\bf G}_{u_1u_2}
{{\id}\over{\id+{\bf B}_{u_2}{\bf G}_{u_2u_2}}}
{\bf B}_{u_2} {\bf G}_{u_2u_1}
\right\}
\end{equation}

\noindent Similarly, they found 
that the bulk, leading-order,
contribution to the three-point probability is
given by

\begin{eqnarray}
\nonumber
\det (\id+{\bf BG}) & = & \\
\nonumber
& &
\hspace{-0.5in}
p_{u_1}p_{u_2}p_{u_3}
{\rm Trace}
\left\{
{{\id}\over{\id+{\bf B}_{u_1}{\bf G}_{u_1u_1}}}
{\bf B}_{u_1} {\bf G}_{u_1u_2}
{{\id}\over{\id+{\bf B}_{u_2}{\bf G}_{u_2u_2}}}
{\bf B}_{u_2} {\bf G}_{u_2u_3}
{{\id}\over{\id+{\bf B}_{u_3}{\bf G}_{u_3u_3}}}
{\bf B}_{u_3} {\bf G}_{u_3u_1}
\right\} + \\
& & 
\nonumber
\hspace{-0.5in}
p_{u_1}p_{u_2}p_{u_3}
{\rm Trace}
\left\{
{{\id}\over{\id+{\bf B}_{u_1}{\bf G}_{u_1u_1}}}
{\bf B}_{u_1} {\bf G}_{u_1u_3}
{{\id}\over{\id+{\bf B}_{u_3}{\bf G}_{u_3u_3}}}
{\bf B}_{u_3} {\bf G}_{u_3u_2}
{{\id}\over{\id+{\bf B}_{u_2}{\bf G}_{u_2u_2}}}
{\bf B}_{u_2} {\bf G}_{u_2u_1}
\right\}  \\
\label{eq:MR.Trace.3}
\end{eqnarray}

\end{widetext}

\noindent (\Maybeeqs{\ref{eq:MR.Trace.2}-\ref{eq:MR.Trace.3}}
are written in a different form than the
expressions in~\cite{Mahieu.Ruelle}, but are equivalent.)

The derivation in~\cite{Mahieu.Ruelle}
of \maybeeq{\ref{eq:MR.Trace.2}}
in the bulk relied on the 
fact that the
leading-order contribution to the two-point
function comes from the pieces of 
$\det (\id+{\bf B}{\bf G})$ with two terms 
off the block diagonal (i.e. one term from
${\bf G}_{u_1u_2}$, and one term from ${\bf G}_{u_2u_1}$).
Similarly, the derivation of \maybeeq{\ref{eq:MR.Trace.3}}
was based on the fact that the
leading-order, connected, contribution to the three-point
function comes from the terms of 
$\det (\id+{\bf B}{\bf G})$ with three terms
off the block diagonal. 

The trace formulae can be extended 
for all
higher-order correlations for the open case.
We will see that the leading-order contribution to the
open boundary $n$-point function decays as
$\mathcal{O}(1/x^{2n})$.
The open boundary Green function 
(appendix~\ref{sec:Green.Functions})
between $(x_1,y_1)$ and $(x_2,y_2)$ is

\begin{equation}
\label{eq:G.open.leading}
G_{\op ,0} (x_1,y_1;x_2,y_2) = 
{{(y_1+1)(y_2+1)} \over  {\pi (x_1-x_2)^2}} +\dots
\end{equation}

\noindent Here, $x$ labels distance along the boundary, and
$y$ labels distance from the boundary (the boundary is
at $y=0$). Since the Green function decays as $1/x^2$, we
can only have $n$ terms off
the block diagonal. Furthermore, to get a connected
correlation function, we must have exactly one
term off the block diagonal in every block row and in
every block column. This allows us to generalize
\maybeeqs{\ref{eq:MR.Trace.2}-\ref{eq:MR.Trace.3}}
for open boundary $n$-point functions; they generalize
in the obvious manner, with $(n-1)!$ trace terms
for the $n$-point function, corresponding to the 
$(n-1)!$ ways
that we can loop through the $n$ positions.

\Maybeeq{\ref{eq:G.open.leading}} 
shows that each off-diagonal block,
${\bf G}_{uv}$, factorizes into the product of
a column vector and row vector:

\begin{equation}
{\bf G}_{u_fu_g} = 
{1\over {\pi (x_f-x_g)^2}} {\bf h}_{u_f} {\bf h}_{u_g}^T \ ,
\end{equation}

\noindent where ${\bf h}_{u_f}$ is a column
vector of heights $y+1$ of the sites
around $x_f$ in $L_{\op,u_f}(x_f)$---i.e. the $p^{th}$
entry of ${\bf h}_{u_f}$ is the value of
$y+1$ for the
$p^{th}$ site of $L_{\op,u_f}(x_f)$.
Substituting this in
the generalization of
\maybeeqs{\ref{eq:MR.Trace.2}-\ref{eq:MR.Trace.3}},
and using the cyclicity of the trace,
each of the $(n-1)!$ matrix traces becomes a product of 
$n$ $1\times 1$ matrices. Furthermore, the 
$(n-1)!$ traces differ from
each other only in the
$1/(x_f-x_g)^2$ terms chosen. The
leading-order, connected part of the correlation function 
of $n$ local arrow diagrams is then found to be

\begin{eqnarray}
\nonumber
\left< L_{\op,u_1}(x_1) L_{\op,u_2}(x_2) \dots L_{\op,u_n}(x_n) \right>
& = & \\
& & \hspace{-1.0in} = \left( \prod_{f=1}^n k_{u_f} \right)
\det {\bf M}
\end{eqnarray}

\noindent $\bf M$ is defined as the $n\times n$ matrix 

\begin{equation}
\label{eq:MMatrix}
M_{fg} \equiv \left\{ 
\begin{array}{cl}
0 & \rm{if}\ f=g \\
1/(x_f-x_g)^2 & \rm{if}\ f\neq g
\end{array}
\right. \,
\end{equation}

\noindent and the $k_u$ are simply numbers:

\begin{equation}
\label{eq:k.from.trace}
k_u \equiv {1\over\pi}
\det \left(\id+{{\bf B}_u{\bf G}_{uu}}\right)
\left( {\bf h}_u^T 
{\id\over{\id+{\bf B}_u{\bf G}_{uu}}}
{\bf B}_u {\bf h}_u \right)
\end{equation}

Inserting this into
\maybeeq{\ref{eq:open.height.to.local}} gives
all open boundary $n$-point correlations. To express
our results in a simpler manner, we define

\begin{equation}
\phi_a(x) \equiv {{\delta_{h_x,a}-p_{a,{\op}}} \over K_a} \qquad,
\rm{where}\ 
a=1,\dots 4
\end{equation}

\noindent  We have defined the following constants:

\begin{eqnarray}
\begin{array}{ll}
p_{1,{\op}}={9\over 2}-{42\over\pi}+
{320\over{3\pi^2}}-{512\over{9\pi^3}} &
K_1=-{3\over\pi}+{80\over{3\pi^2}}-{512\over{9\pi^3}} \\ \\
p_{2,{\op}}=-{33\over 4}+{66\over\pi}- 
{160\over\pi^2}+{1024\over{9\pi^3}} &
K_2={9\over\pi}-{200\over{3\pi^2}}+{1024\over{9\pi^3}} \\ \\
p_{3,{\op}}={15\over 4}-{22\over\pi}+
{160\over{3\pi^2}}-{512\over{9\pi^3}} & 
K_3=-{7\over\pi}+{40\over\pi^2}-{512\over{9\pi^3}} \\ \\
p_{4,{\op}}=1-{2\over\pi} & 
K_4={1\over\pi}
\end{array}
\end{eqnarray}

\noindent $p_{a,{\op}}$ is the probability for a site along 
an open boundary to have height $a$,
and the $K_a$ are normalization
factors. We then, finally, have

\begin{equation}
\label{eq:nptcorr}
\left<\phi_{a_1}(x_1)\phi_{a_2}(x_2)\dots\phi_{a_n}(x_n)\right> 
=\det({\bf M})
\end{equation}

\noindent For $n=2$, this reproduces the
open boundary
one- and two-point functions 
found in~\cite{Ivashkevich}.

$\det ({\bf M})$ is the same as the $n$-point
function of $-2\lcftbdy$, so
up to rescaling factors ($-2K_a$'s),
all four height variables are represented
by $\lcftbdy$ along open boundaries.
This is rather surprising, given that we have
seen that the height variables are represented
by different operators along closed boundaries
(\maybeeqs{\ref{eq:height.one.identification}-\ref{eq:height.three.identification}}).
In CFT's, boundary operators are derived from
OPE's of bulk operators---so the fact that the
height operators are different along closed
boundaries proves that they must be different
in the bulk, but apparently these different
bulk operators become identical
along open boundaries.

We nowhere used the fact that these were the local
arrow diagrams associated with the height variables.
So, in fact, we have shown that {\it all} 
local arrow diagrams
along open boundaries are represented by
$\partial\theta\partial\bar\theta$.

We have also found the correlation function of $n$ unit
height variables along {\it closed} boundaries.
This requires local arrow constraints at $3n$ 
vertices of the ASM, and thus the calculation of a
$3n$-dimensional matrix determinant. The matrix is divided 
into 3 by 3 block submatrices, such that the diagonal blocks 
are all identical, and the off-diagonal blocks all have 
the same form. A rotation makes the matrix 
diagonal in 2 out of every 3 rows (and columns).  The universal 
part of the correlation function is 
thus found to be 

\begin{equation}
\label{eq:unit.closed.corr}
\left({{3\pi-8}\over\pi^2}\right)^n 
{\rm det} {\bf M} 
\end{equation}

\noindent This confirms the field identification in
\maybeeq{\ref{eq:height.one.identification}}.


\section{\protect\boldmath$n$-POINT CORRELATIONS ALONG
OPEN BOUNDARIES, WITH DISSIPATIVE DEFECTS}
\label{sec:all.open.correlations.dissipation}

Along open boundaries, the
defect-free Green function, 
${\bf G}_0={\bf G}_{\op,0}$ does not diverge
as $L\to\infty$, so for a single dissipative
defect we can modify the Green function as in
\maybeeq{\ref{eq:defect.Green}}.
Using this new Green function, the open height
probabilities at $(x_1,0)$, for a defect 
of strength $k$ at $\vec{d}=(0,y)$ are

\begin{equation}
\label{eq:open.single.defect}
{\rm f}_\op (a)= -K_a
{{k(y+1)^2} \over {\pi (1+k G_{\op,0}(\vec{d},\vec{d}))}}
{1 \over {x_1^4}}, \ \ \ \ a=1,2,3,4
\end{equation}

\noindent The same $K_a$ factors that we saw
in the height-height correlations appear in height-defect
correlations.

We define an operator
$\phi_{5;k} (\vec{d})$,
corresponding to the addition of a 
defect of strength $k$ at $\vec{d}=(x,y)$,
and then multiplication 
of all correlation functions by a normalization
factor 

\begin{equation}
\label{eq:defect.normalization}
{\pi (1+k G_{\op,0}(\vec{d},\vec{d}))}\over{k(y+1)^2}
\end{equation}

\noindent Then \maybeeq{\ref{eq:open.single.defect}}
becomes

\begin{equation}
\label{eq:open.defect.2pt}
\left< \phi_a (x_1) \phi_{5;k} (x_2) \right> =
- {1 \over {(x_1-x_2)^4}},\ \ \ a\neq 5
\end{equation}

\noindent $\phi_{5;k}$ acts just like 
any of the four height variables in two-point
correlations (\maybeeq{\ref{eq:open.defect.2pt}} is
\maybeeq{\ref{eq:nptcorr}} with $n=2$).
In fact, we find that $\phi_{5;k}$
acts like
$\phi_1$, $\phi_2$, $\phi_3$, and $\phi_4$ in all
higher-order
correlation functions, containing multiple height
variables and
multiple dissipative defects. 

Suppose we are calculating a correlation function with
$n$ height variables, and $m$ dissipative defects.
The dissipative defects
are at $\vec{d}_w=(x_w,y_w)$, and have strength $k_w$,
$1 \leq w \leq m$. 
As with the height locations, the $x_w$ coordinates of the
defects all scale with the same factor $x$, where
$x\to\infty$.
The change in the toppling
matrix , $\delta{\bf\Delta}\equiv{\bf\Delta}-{\bf\Delta}_0$, is

\begin{equation}
\delta\Delta_{\vec{i},\vec{j}} = \left\{
\begin{array}{cl}
k_w & \rm{if}\ \vec{i}=\vec{j}=\vec{d}_w,\quad 1\leq w \leq m \\
0 & \rm{otherwise}
\end{array}
\right.
\end{equation}

\noindent The Green function is modified from its 
defect-free value, ${\bf G}_{\op,0}$, to

\begin{eqnarray}
\nonumber
{\bf G} & = &
{\id\over\bf\Delta}=
{\id\over{{\bf\Delta}_0+\delta{\bf\Delta}}} = 
{{{\bf G}_{\op,0}}\over{\id+(\delta{\bf\Delta})
{\bf G}_{\op,0}}} \\
& = &
\sum_{p=0}^\infty {\bf G}_{\op,0}
\left( - (\delta{\bf\Delta}) {\bf G}_{\op,0} \right)^p
\label{eq:Green.multiple.defect}
\end{eqnarray}

\noindent $G(\vi ,\vj )$ can be represented 
as a trip from $\vec{i}$ to $\vec{j}$, where
along the trip, the traveller can visit any of the defect
sites as often as he or she wishes, each time 
picking up a factor of $-(\delta {\bf\Delta}) {\bf G}_{\op,0}$.

We have already seen that the 
defect-free correlation function of $n$ height variables
has a leading term of 
$\mathcal{O}(1/x^{2n})$. If we instead use the Green function with
defects, each trip to a defect introduces a factor of
$1/x^2$ (see \maybeeq{\ref{eq:G.open.leading}}).
In a connected function, we should visit each defect at
least once; in the leading term, each defect
will be visited from a distant site exactly once, and
the correlation function
will have a leading term of 
$\mathcal{O}(1/x^{2(n+m)})$.

After visiting $\vec{d}_w$, we may travel repeatedly from
$\vec{d}_w$ to $\vec{d}_w$ without picking up extra
factors of $1/x^2$.
This produces a contribution to
\maybeeq{\ref{eq:Green.multiple.defect}} of

\begin{equation}
\label{eq:defect.geometric.sum}
\sum_{p=0}^\infty 
(-k_w G_0 (\vec{d}_w,\vec{d}_w))^p = 
{1 \over{1+k_w G_0 (\vec{d}_w,\vec{d}_w)}}
\end{equation}

\noindent (We already saw this factor
for a single dissipative defect, in
\maybeeq{\ref{eq:defect.Green}}.)
Furthermore, inspection of
\maybeeq{\ref{eq:G.open.leading}}
shows that the visit to the defect at $\vec{d}_w$
from another site will result in a factor of
$k_w(y_w +1)^2/\pi$. With
\maybeeq{\ref{eq:defect.geometric.sum}}, 
this motivates the normalization
factor in \maybeeq{\ref{eq:defect.normalization}}.

\Maybeeq{\ref{eq:defect.normalization}}
normalizes the correlation function of $n$
height variables and $m$ defects. To see that 
the form of the correlation function is
still $\det ({\bf M})$, expand the determinant out into
cycles. The connected part of the 
determinant in \maybeeq{\ref{eq:nptcorr}}
is a sum of closed cycles of length $n$, where
each cycle visits
each of the positions
$(x_f,0)$ exactly once, and picks up a factor of
$1/(x_f-x_g)^2$ when it travels from
$x_f$ to $x_g$. After normalizing, the defects 
have exactly the same effect as the height 
variables---each trip
to (or from) a defect results in a $1/x^2$ term
from the Green function to (or from) the
defect (\maybeeq{\ref{eq:G.open.leading}}).

So, in the end, the correlation function of $n$
height variables on the boundary and $m$ defect sites 
near or on the boundary, is given by the 
$(m+n)$ dimensional matrix determinant, 
$\det ({\bf M})$ (with appropriate normalization factors).
This shows that dissipative defect sites
along or near open boundaries
are, like the height variables, represented by
$\lcftbdy$.

Note that a dissipative defect has a much larger effect 
along a closed boundary than along an open one.
A defect is represented by a 
dimension zero operator along a closed boundary,
but by a dimension two operator along an  open boundary.
This makes sense; along open boundaries, 
grains are already dissipated by
topplings, so adding a little more dissipation 
has only minor effects,
compared to dissipation on a closed boundary.


\acknowledgments{After 
the bulk of this work was completed, we were
informed of independent, then unpublished,
calculations of 
\maybeeqs{\ref{eq:f111}-\ref{eq:f122}}, in the
massive case, by G. Piroux and P. 
Ruelle~\cite{PR.parallel}.
Discussions with G. Piroux and P. Ruelle
then led us to the field identifications in
\maybeeqs{\ref{eq:height.two.identification}-\ref{eq:height.three.identification}},
and they further corrected an error we had made in these
same
equations.
This work was supported by
Southern Illinois University Edwardsville.
We thank V. Gurarie and E. V. Ivashkevich 
for useful discussions.}


\appendix


\section{ANOMALOUS GRAPHS IN
BOUNDARY TWO-POINT CORRELATIONS---PART I}
\label{sec:Complication1}

In this section we discuss what we call
anomalous graphs of the first kind, which arise when
converting from ASM height probabilities to spanning tree
probabilities. As stated in 
section~\ref{sec:Closed.Loop.Calculation},
it would be natural to expect, based on analogy with the
one-point height probabilities, for the
two-point height probabilities to
be given by \maybeeq{\ref{eq:Nonlocal.To.Height.Two}}.
However, this turns out to not be the case. Let us carefully
consider how height correlations can be turned into spanning
tree probabilities. We focus on the closed two-point
correlations; other cases are similar.

\begin{figure}[tb]
\epsfig{figure=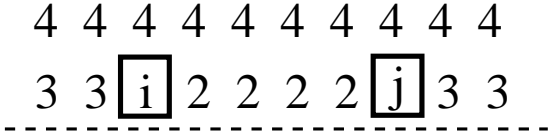,width=3.0in}
\caption{State not in any $S_{kl}$, and in multiple
${\tilde S}_{kl}$.}
\label{fig:FSC.Complications}
\end{figure}

For correlations between $\vi$ and $\vj$, 
Ivashkevich divided the states 
of the ASM into sets $S_{kl}$, 
consisting of states
allowed when $h_{\vi }\geq k$ and $h_{\vj }\geq l$, but forbidden
otherwise~\cite{Ivashkevich}. 
However, not all ASM states fall into one of these 
sets. There are states that are allowed when 
$\htpair{1}{2}$, and when $\htpair{2}{1}$,
but forbidden when $\htpair{1}{1}$; these states
do not belong to any set $S_{kl}$. One 
such state is shown in 
figure~\ref{fig:FSC.Complications}.

\begin{figure}[tb]
\epsfig{figure=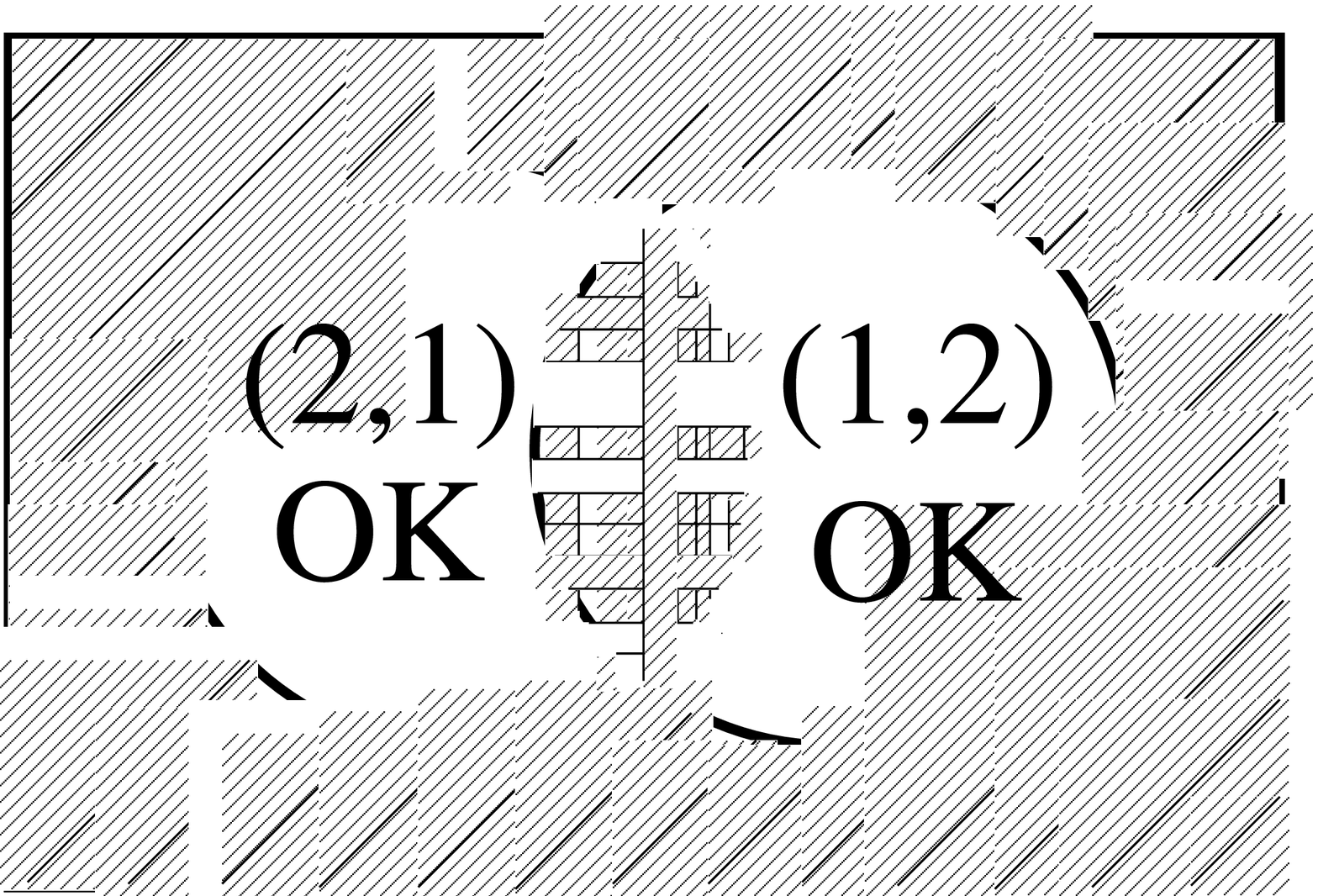,width=3.3in}
\caption{Venn diagram in the space of states where
$\htpair{2}{2}$ is allowed, but $\htpair{1}{1}$ is not.}
\label{fig:Venn.Diagram}
\end{figure}

We find it convenient to define $\tilde{S}_{kl}$,
consisting of ASM 
height configurations on the sandpile,
{\it excluding} $\vi$ and $\vj$, that are allowed
when we add $\htpair{k}{l}$, but both forbidden when
we add
$\htpair{k-1}{l}$, and also forbidden when 
we add $\htpair{k}{l-1}$.
Note that when counting states, the
fact that we do not specify the heights
of $\vi$ and $\vj$
introduces a multiplicative factor;
for example, 
$\mid\tilde{S}_{12}\mid=\mid S_{12}\mid/(3*2)$ for
$\vi$ and $\vj$ both on a
closed boundary.
Now, every state must be in at least one of the sets
$\tilde{S}_{kl}$, but
some states are in several $\tilde{S}_{kl}$'s.
For example, the state in
figure~\ref{fig:FSC.Complications} is in both
$\tilde{S}_{12}$ and $\tilde{S}_{21}$.

No anomalous graphs arise for two-point correlations involving at 
least one unit height variable,  since in those cases
the relevant $\tilde{S}$-sets do not intersect. 
The number of states where $\vi$ has
height one and $\vj$ has height $h$ is 
$\sum_{p=1}^h \mid {\tilde S}_{1p} \mid$.
Furthermore, the
representation of $\tilde{S}_{1h}$ is exactly
what we would expect;
it corresponds to the number of spanning trees where
$\NNP_{\vi}=0$, and $\NNP_{\vj}=h-1$. 
So no anomalous graphs arise when the two-point correlation
has at least one unit height variable.

Things get more complicated when 
both heights are higher heights.
We discuss in detail the 2-2 (height two-height two)
correlation
along a closed boundary; the
analysis for the other two-point correlations is similar.

If $\vi$ and $\vj$ both have height two, we must be in at
least one $\tilde{S}_{kl}$, for $k\leq 2$, $l\leq 2$.
$\tilde{S}_{12}$ and
$\tilde{S}_{21}$ intersect, so 
number of 2-2 states is

\begin{equation}
\mid\tilde{S}_{11}\mid +
\mid\tilde{S}_{12}\mid +
\mid\tilde{S}_{21}\mid +
\left( \mid\tilde{S}_{22}\mid - 
\mid \tilde{S}_{12} \cap \tilde{S}_{21} \mid\right)
\end{equation}

The first three terms
all have the ``natural'' spanning tree representation.
The difficulty is in evaluating
$\mid \tilde{S}_{22} \mid -
\mid \tilde{S}_{12} \cap \tilde{S}_{21} \mid$.
$\tilde{S}_{22}$ counts states where
$\htpair{2}{2}$ is allowed, but neither
$\htpair{1}{2}$ nor $\htpair{2}{1}$ are allowed.
In figure~\ref{fig:Venn.Diagram},
we have started with a large rectangle,
representing the
the set of states
where $\htpair{2}{2}$ is allowed, and
$\htpair{1}{1}$ is forbidden---we
call this set $X$. In the rectangle are
two subsets,
corresponding to
regions where $\htpair{1}{2}$
is allowed, and where $\htpair{2}{1}$
is allowed---we
call these two sets $X_1$ and $X_2$. In this Venn
diagram, $\tilde{S}_{22}$ is the diagonally shaded
region outside the circles, and
$\tilde{S}_{12} \cap \tilde{S}_{21}$
is the cross-hatched intersection of the two circles.
Looking at the Venn diagram, we see that
to find 
$\mid \tilde{S}_{22} \mid -
\mid \tilde{S}_{12} \cap \tilde{S}_{21} \mid$,
we start with the set $X$, and then subtract off
the states in $X_1$ and $X_2$ {\it independently}.
By independently we mean that states in the
intersection of $X_1$ and $X_2$ 
get subtracted off twice. So

\begin{equation}
\mid \tilde{S}_{22} \mid -
\mid \tilde{S}_{12} \cap \tilde{S}_{21} \mid=
\mid X \mid - \mid X_1 \mid - \mid X_2 \mid
\end{equation}

For states in $X$, setting $\htpair{1}{1}$
produces an
MFSC. (Note that we have defined the MFSC as the
maximal FSC produced when the heights
at both $\vi$ and $\vj$ are simultaneously set to
one; if only one height was set to one,
then the largest FSC might be smaller, or there
might be no FSC at all.)
The set $X$ can be partitioned into the following disjoint
subsets, depending on the shape of the MFSC:

\begin{widetext}

\begin{eqnarray}
\nonumber
X_A &:& \rm The\ MFSC\ consists\ of\ disjoint\ subsets\
around\ \vi\ and\ \vj \\
\nonumber
X_B &:& \rm The\ MFSC\ is\ connected,\ with\ 
\geq 2\ neighbors\ of\ \vi\ and\
exactly\ 1\ neighbor\ of\ \vj \\
\nonumber
X_C &:& \rm The\ MFSC\ is\ connected,\ with\ 
\geq 2\ neighbors\ of\ \vj\ and\
exactly\ 1\ neighbor\ of\ \vi \\
\nonumber
X_D &:& \rm The\ MFSC\ is\ connected,\ with\ 
exactly\ 1\ neighbor\ of\ \vi\ and\
exactly\ 1\ neighbor\ of\ \vj 
\end{eqnarray}

\end{widetext}

\noindent $X_1$ and $X_2$ can be partitioned into analogously
defined subsets, $X_{1A}$, $X_{1B}$, etc. . .
So, for example, $X_{1D}$ is the subset of 
$X_D$ such that $\htpair{1}{2}$ is allowed.
(Note that $X_{1C}=\emptyset$ and $X_{2B}=\emptyset$.)
We want

\begin{equation}
\label{eq:X.decomposition}
\sum_{k\in\{A,B,C,D\}}
\left( \mid X_k \mid - \mid X_{1k} \mid - \mid X_{2k} \mid
\right)
\end{equation}

\noindent Our ``naive'' guess would be that this would equal
the set of states where $\NNP_{\vi}=\NNP_{\vj}=1$.
We carefully count the states, comparing with this guess.

$X_A$, after subtracting off the states from
$X_{1A}$ and $X_{2A}$, is equal to one-fourth the
number of spanning trees for which $\vi$ and $\vj$ each
have one $\NNP$, and $\{\vi\} \cup \tree_{\vi}$
does not border 
or intersect $\{\vj\}\cup\tree_{\vj}$.
(The one-fourth comes from the fact that the 
spanning tree arrows
from $\vi$ and $\vj$ can point out from the MFSC in
any direction.)
$\tree_{\vi}$ refers to the set of sites that are
predecessors of $i$.
The condition that
$\{\vi\}\cup\tree_{\vi}$ and 
$\{\vj\}\cup\tree_{\vj}$ cannot border each other
comes from the condition that the MFSC consist of disjoint
subsets around $\vi$ and $\vj$.

We now consider $X_B$. The MFSC generated when
$\htpair{1}{1}$ must still be an MFSC when
$\htpair{2}{1}$.
So $\htpair{2}{2}\to (2,1)$
produces an MFSC that includes exactly one neighbor of $\vj$,
the site
$\vi$, and at least two of $\vi$'s neighbors.
Just as in section~\ref{sec:Height.Probabilities},
this is equivalent
to a modified ASM, where bonds
along the border of the MFSC are removed
(except for one bond of $\vj$).
In \maybeeq{\ref{eq:X.decomposition}},
$X_{2B}=\emptyset$, but we do
need to subtract off states in $X_{1B}$. 
To do this we 
only count the states of $X_B$ such that
$\htpair{1}{2}$ is forbidden, which implies that
$h_{\vi}=2\rightarrow 1$ should produce a new, smaller FSC,
completely contained within the 
larger MFSC.
We then see that 
$\mid X_B\mid-\mid X_{1B}\mid -\mid X_{2B}\mid$
corresponds to one-fourth of all spanning trees
where $\NNP_{\vi}=\NNP_{\vj}=1$, and either
$\vi\in\tree_{\vj}$
or $\vi$ borders $\tree_{\vj}$, {\it with one exception}.
The exception occurs
because the MFSC, by construction of $X_B$,
can only have one neighbor of $\vj$.
So $\mid X_B\mid - \mid X_{1B}\mid - \mid X_{2B}\mid$
will not count
cases where $\NNP_{\vi}=NNP_{\vj}=1$,
$\vi$ borders $\tree_{\vj}$, and $\vj$ borders 
$\tree_{\vi}$---this case is shown in 
figure~\ref{fig:AnomalousGraphC}.
We label this set of graphs as $c$. Since $c$ has
$\NNP_{\vi}=\NNP_{\vj}=1$, it would be natural to
expect it to appear 
in the spanning trees contributing
to the height two-height two correlation function.
However, since no MFSC's of $X$ have two
neighbors of $\vi$ and two neighbors of $\vj$,
graph $c$ 
does not appear in $X_C$ or $X_D$
either.

The analysis for $X_C$ is identical to that
for $X_B$, and \linebreak
$\mid X_C\mid - \mid X_{1C}\mid - \mid X_{2C}\mid$ counts
one-fourth the spanning trees
where $\NNP_{\vi}=\NNP_{\vj}=1$, and either 
$\vj\in\tree_{\vi}$
or $\vj$ borders $\tree_{\vi}$, except that, again,
the spanning trees of $c$ are excluded.

In $X_D$, the MFSC has only one neighbor of $\vi$ 
and one neighbor of $\vj$.
The one-to-one mapping between $X_D$ and spanning tree
states posessess some subtleties,
but the end result is what 
one would expect: The number of
states in 
$\mid X_D\mid - \mid X_{1D}\mid - \mid X_{2D}\mid$ 
is one-fourth the number of spanning
trees where $\NNP_{\vi}=\NNP_{\vj}=1$, and 
$\tree_{\vi}$ and $\tree_{\vj}$ border each other, but 
$\vi \notin \tree_{\vj}$, and $\vj \notin\tree_{\vi}$.

In the end, we see that
$4\left(\mid\tilde{S}_{22}\mid -
\mid\tilde{S}_{12}\cap\tilde{S}_{21}\mid\right)$
is equal to the number of spanning trees
where $\NNP_{\vi}=\NNP_{\vj}=1$,
except for 
the set $c$, which contains all spanning trees
where
both $\vi$ borders $\tree_{\vj}$, and 
$\vj$ borders $\tree_{\vi}$.
$c$ consists of the
anomalous graphs of 
the first kind.

An similar analysis  for other closed correlation
functions shows that the spanning trees in
$c$ contribute to the height two-height three
correlation, and get subtracted from the 
height three-height correlation
(relative to a ``naive'' approach).
These changes are necessary for
the height probabilities to all sum to one, so this provides
a check on our mapping between ASM states and spanning tree
states.
In the next appendix, we
consider yet another complication
that arises in the calculation of the two-point functions.


\section{ANOMALOUS GRAPHS IN BOUNDARY TWO-POINT CORRELATIONS---PART II}
\label{sec:Complication2}

We saw in the previous appendix that
\maybeeq{\ref{eq:Nonlocal.To.Height.Two}}
does not quite hold, but is only off by the
anomalous graph $c$.
So except for this complication,
the two-point height correlations can be turned into linear
combinations of probabilities for spanning trees
with nonlocal conditions on $\NNP_{\vi}$ and $\NNP_{\vj}$.
As in the previous appendix, we discuss only the
closed boundary two-point functions; other cases are 
similar.

\begin{figure}[tb]
\epsfig{figure=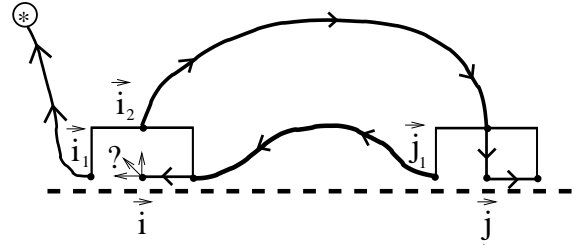,width=3.0in}
\caption{Correlations where $\beta_1\neq\beta_2$}
\label{fig:beta1.not.beta2}
\end{figure}

As discussed in section~\ref{sec:Ivashkevich.Review}, we
can write the probability 
$\N_h$ to have $\NNP_{\vi}=h-1$ as
a linear combination of nonlocal arrow diagrams, 
which we can then rewrite
as follows (see figures~\ref{fig:NonlocalList} 
and~\ref{fig:LocalList}):

\begin{eqnarray}
\nonumber
\N_2 & = & 2 (\beta_1+\beta_2+\gamma) \\
\nonumber
& = & 2(-L_{c,1}+3L_{c,2}+L_{c,3}) + \\
& & \qquad 4(\beta_1-\beta_2)+2(\phi_1-\phi_2) 
\label{eq:N2.LocalPlus} \\
\nonumber
\N_3 & = & 2 (\phi_1+\phi_2+\epsilon)+\delta \\
\nonumber
& = & 2 (2L_{c,1}-2L_{c,2}+L_{c,5})+L_{c,4}- \\
& & \qquad 4(\beta_1-\beta_2)-2(\phi_1-\phi_2) 
\label{eq:N3.LocalPlus}
\end{eqnarray}

\noindent These relationships hold regardless of the
correlation functions that $\vi$ are in.
We define operators corresponding to the
{\it local} parts of these terms:

\begin{eqnarray}
L_{N2} & \equiv & 2 (-L_{c,1}+3 L_{c,2}+L_{c,3}) \\
L_{N3} & \equiv & 2 (2L_{c,1}-2L_{c,2}+L_{c,5})+L_{c,4}
\end{eqnarray}

\noindent In one-point functions, 
$\beta_1=\beta_2$, and $\phi_1=\phi_2$, so 
$\N_2$ and $\N_3$ simply become
$L_{N2}$ and $L_{N3}$, which are local, and whose
expectation values can be found with the
Majumdar-Dhar method,
as discussed in section~\ref{sec:Ivashkevich.Review}.
(These are the same relations found in~\cite{Ivashkevich}.)

It would be simplest if in correlations of
$\N_2$ and $\N_3$, we could replace $\N_2$ and $\N_3$
with $L_{N2}$ and $L_{N3}$, since local
correlation functions are easily calclated.
As in the previous
appendix, we call this approach
``naive,''---this 
naive approach does not quite work,
and we call the deviations of the correct answers from the
naive approach ``anomalous graphs of the second kind.''

We no longer have
$\beta_1=\beta_2$ and $\phi_1=\phi_2$ in
correlations of $\N_2$ and $\N_3$, 
because in correlations between 
distant sites $\vi$ and $\vj$,
switching arrows at $\vi$ can
affect predecessor relationships at $\vj$. 
To analyze the $(\beta_1-\beta_2)$ terms in
\maybeeqs{\ref{eq:N2.LocalPlus}-\ref{eq:N3.LocalPlus}},
consider the configuration in 
figure~\ref{fig:beta1.not.beta2},
where we have
not specified the direction of the 
arrow from $\vi$. If the arrow 
from $\vi$ points to $\vi_1$, then 
$\vj_1$ is not a predecessor of $\vj$, so 
the configuration at $\vj$ 
is $\gamma$.
And if the arrow from $\vi$ points to $\vi_2$, then
$\vj_1$ is a predecessor of $\vj$, so
the configuration at $\vj$ is $\phi_2$. So
switching from $\beta_1$ to
$\beta_2$ at $\vi$
can affect whether the configuration at $\vj$ is
$\phi_2$ or $\gamma$.
However, this inequivalence between $\beta_1$ and $\beta_2$
turns out to have no effect on any correlation functions,
to any order, since $\phi_2$ and $\gamma$ always
appear in the combination $\phi_2+\gamma$, 
in $L_{N2}$ and $L_{N3}$,
and $\phi_2+\gamma$
has no correlations with $\beta_1-\beta_2$.

\begin{widetext}

\begin{figure}[tb]
\epsfig{figure=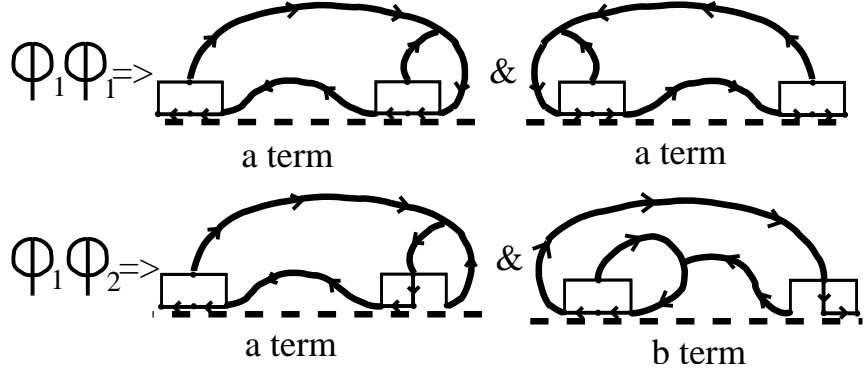,width=4.5in}
\caption{Correlations where $\phi_1\neq\phi_2$}
\label{fig:Phi1.Phi2.Complications}
\end{figure}

\end{widetext}

Unfortunately, things become more complicated
with the $\phi_1$ and $\phi_2$ terms. 
If the long path of $\phi_1$ avoids arrow
restrictions at other sites, the 
long path can be reversed, and
$\phi_1$ will be equivalent to $\phi_2$.
However, if the long path goes through arrow
restrictions at other sites of the correlation
function, then
$\phi_1$ will not be equivalent to $\phi_2$. 
Figure~\ref{fig:Phi1.Phi2.Complications}
compares diagrams that arise in
$(\phi_1,\phi_1)$ correlations, with
diagrams that arise in 
$(\phi_2,\phi_2)$ correlations.
Three of the four
diagrams shown (labeled with ``a'''s) are equivalent,
but the fourth one (labeled with ``b'') is not.
The resultant anomalous graphs of the second kind
were shown and discussed earlier,
in section~\ref{sec:Closed.Loop.Calculation}
(see figure~\ref{fig:AnomalousGraphsAB}).
When all the correlations of the $(\phi_1-\phi_2)$'s in
\maybeeqs{\ref{eq:N2.LocalPlus}-\ref{eq:N3.LocalPlus}}
are considered, we find the following
for closed boundary correlations:

\begin{eqnarray}
\label{eq:ab.contribution.1}
\left<\N_2(r)\N_2(0)\right> & = & 
\langle L_{N2}(r)L_{N2}(0)\rangle +4 (a-b) \\
\label{eq:ab.contribution.2}
\left<\N_2(r)\N_3(0)\right> & = & 
\langle L_{N2}(r)L_{N3}(0)\rangle -4 (a-b) \\
\label{eq:ab.contribution.3}
\left<\N_3(r)\N_3(0)\right> & = & 
\langle L_{N3}(r)L_{N3}(0)\rangle +4 (a-b) 
\end{eqnarray}

\noindent The correlations involving the height one variable
are unaffected by these complications.

To summarize, naively transforming from nonlocal arrow
diagrams to local arrow diagrams {\it independently}
at every site of a two-point correlation
results in anomalous graphs $a$ and $b$. 


\section{GREEN FUNCTIONS}
\label{sec:Green.Functions}

The inverse of the bulk toppling matrix 
${\bf \Delta}_0$ is the lattice Green function, which has
long been known~\cite{Spitzer}. It is given by

\begin{equation}
G_0 (\tilde{x},\tilde{y}) = 
\int_0^{2\pi} {{dp_1}\over{2\pi}} 
\int_0^{2\pi} {{dp_2}\over{2\pi}} 
{{e^{ip_1\tilde{x}+ip_2\tilde{y}}}\over{4-2\cos p_1-2\cos p_2}}
\end{equation}

\noindent This integral is divergent, producing terms of 
order $\ln L$, where $L$ is the system size, but these
divergences are usually unimportant, since we are typically
concerned with differences in Green functions.
For large $\tilde{x}$, $\tilde{y}$ this has the 
expansion~\cite{Spitzer}

\begin{equation}
G_0 (\tilde{x},\tilde{y}) = - {1\over{4\pi}} 
\ln (\tilde{x}^2+\tilde{y}^2) 
-{\gamma\over{2\pi}}
-{{\ln 8}\over{4\pi}}+\dots \ ,
\end{equation}

\noindent where $\gamma =0.57721\dots$ is the Euler-Mascheroni 
constant. For sites $(x_1,y_1)$ and $(x_2,y_2)$
near an open boundary, where $x$ is the coordinate along the
boundary, and $y$ is the distance from the boundary
(located at $y=0$), the Green function is~\cite{Bdy.Falloff}

\begin{eqnarray}
\nonumber
G_{\op,0} (x_1,y_1;x_2,y_2) & = & G_0(x_1-x_2,y_1-y_2) - \\
& & \hspace{-0.2in} G_0(x_1-x_2,y_1+y_2+2)
\label{eq:Green.open.general}
\end{eqnarray}

\noindent Along closed boundaries, it is~\cite{Bdy.Falloff}

\begin{eqnarray}
\nonumber
G_{\cc,0} (x_1,y_1;x_2,y_2) & = & G_0(x_1-x_2,y_1-y_2) + \\
& & \hspace{-0.2in} G_0(x_1-x_2,y_1+y_2+1)
\label{eq:Green.closed.general}
\end{eqnarray}

\noindent The minus sign between the Green functions in
\maybeeq{\ref{eq:Green.open.general}} 
cancels out divergences in the Green function.
The expansion of the 
Green function for points
along the boundary has already been
calculated~\cite{Bdy.Falloff}, and can be extended to
points near the boundary, but far from each
other
($y_1=\mathcal{O}(1)$, $y_2=\mathcal{O}(1)$,
$\mid x_1-x_2\mid \to\infty$),
by the recursion relationships,
${\bf G}_0{\bf\Delta}_0=\id$. We find

\begin{equation}
\label{eq:Green.open.r}
G_\op (x_1,y_1;x_2,y_2)=
{{(y_1+1)(y_2+1)}\over{\pi (x_1-x_2)^2}} +\dots
\end{equation}

\noindent and

\begin{widetext}

\begin{eqnarray}
\nonumber
G_\cc (x_1,y_1,x_2,y_2) & = &
-{1\over\pi} \ln \mid x_1-x_2 \mid
-({\gamma\over\pi}+{{3\ln 2}\over{2\pi}}) 
- (3y_1(y_1+1)+3y_2(y_2+1)+1)
{1\over{6\pi (x_1-x_2)^2}} \\
& & \hspace{-1.0in}
+\left( 
y_1(y_1+1)(y_1^2+y_1-1)+
y_2(y_2+1)(y_2^2+y_2-1)  
+6 y_1(y_1+1)y_2(y_2+1)-{17\over 60}
\right)
{1\over{4\pi (x_1-x_2)^4}}+\dots
\label{eq:Green.closed.r}
\end{eqnarray}

\end{widetext}


\section{MORE CLOSED BOUNDARY CORRELATION FUNCTIONS}
\label{sec:Additional3pts}

Here we list the three-point correlation functions along
closed boundaries that were not stated in
section~\ref{sec:Three.Point.Closed}.
As a check, the correlation functions in this appendix
were found by the methods already described in 
sections~\ref{sec:Closed.Loop.Calculation}
and~\ref{sec:Three.Point.Closed}.
However, they can all also be determined from those
already listed in 
section~\ref{sec:Three.Point.Closed},
from the requirement that
the three height probabilities must sum 
at all sites, and by symmetry. They are listed here only
for reference, and because they provide checks on
our calculations. We find

\begin{eqnarray}
\nonumber
f_\cc (1,1,3) & = & - {{2(4-\pi)(3\pi-8)^2} \over
{\pi^6 (x_1-x_2)^2(x_1-x_3)^2(x_2-x_3)^2}} \\
& & + {{(3\pi-8)^2} \over {\pi^5
(x_1-x_3)^3(x_2-x_3)^3}}+\dots \\
\nonumber
f_\cc (1,2,3) & = & {{4(\pi-3)(\pi+8)(3\pi-8)} \over
{\pi^6 (x_1-x_2)^2(x_1-x_3)^2(x_2-x_3)^2}} \\
\nonumber
& & - {{(3\pi-8)^2} \over {\pi^5
(x_1-x_2)^3(x_2-x_3)^3}}+\\
& & - {{(3\pi-8)(24-7\pi)}\over
{2\pi^5 (x_1-x_2)^3(x_1-x_3)^3}} \\
\nonumber
f_\cc (1,3,3) & = & {{(3\pi-8)(4-\pi)(8+\pi)} \over
{\pi^6 (x_1-x_2)^2(x_1-x_3)^2(x_2-x_3)^2}} \\
& & + {{(3\pi-8)(8-\pi)} \over {2\pi^5 (x_1-x_2)^3(x_1-x_3)^3}}
+\dots
\end{eqnarray}

\noindent We can now check that 
$f_c(1,1,1)+f_c(1,1,2)+f_c(1,1,3)=0$, as it must.
Interchanging $x_2$ and $x_3$ in $f_c(1,1,2)$ gives

\begin{eqnarray}
\nonumber
f_\cc (1,2,1) & = & - {{8(\pi-3)(3\pi-8)^2} \over
{\pi^6 (x_1-x_2)^2(x_1-x_3)^2(x_2-x_3)^2}} \\
& & - {{(3\pi-8)^2} \over 
{\pi^5 (x_1-x_2)^3(x_3-x_2)^3}}+\dots 
\end{eqnarray}

\noindent We can then check that 
$f_c(1,1,1)+f_c(1,2,1)+f_c(1,3,1)=0$. 
Three-point correlation functions with no unit height
variables {\it cannot} be found with the methods in
this paper, as already discussed in 
section~\ref{sec:Three.Point.Closed}.
However, if we use the conjecture proposed in
section~\ref{sec:Closed.Loop.Calculation},
of dropping the anomalous graphs
(as we did to obtain \maybeeq{\ref{eq:f222}}),
we now obtain

\begin{eqnarray}
\nonumber
f_\cc (2,2,3) & = & {{(24-5\pi)(-192+112\pi-13\pi^2)} \over
{4\pi^6 (x_1-x_2)^2(x_1-x_3)^2(x_2-x_3)^2}} \\
& & + {{(3\pi-8)(7\pi-24)} \over 
{2\pi^5 (x_1-x_3)^3(x_2-x_3)^3}}+\dots \\
\nonumber
f_\cc (2,3,3) & = & - {{(8+\pi)(-192+112\pi-13\pi^2)} \over
{4\pi^6 (x_1-x_2)^2(x_1-x_3)^2(x_2-x_3)^2}} \\
& & - {{(3\pi-8)(8-\pi)} \over 
{2\pi^5 (x_1-x_2)^3(x_1-x_3)^3}}+\dots \\
\nonumber
f_\cc (3,3,3) & =  & -  {{(8+\pi)(64-32\pi+\pi^2)} \over
{4\pi^6(x_1-x_2)^2(x_1-x_3)^2(x_2-x_3)^2}} +\dots, \\
\end{eqnarray}

\noindent As with the other correlation functions in this
section, these agree with the requirements that
the three height probabilities must sum to one at
all sites, and with the field identifications in
\maybeeqs{\ref{eq:height.one.identification}-\ref{eq:height.three.identification}}.


\section{THE \protect\boldmath$c=-2$ CONFORMAL FIELD THEORY}
\label{sec:c-2LCFT}

The central charge -2 conformal field theory is
perhaps the simplest known logarithmic conformal
field theory. While the theory has a simple
underlying Gaussian structure, it 
still possesses a number of subtle features. 
We use the formulation of the $c=-2$ CFT where
the action is given by

\begin{equation}
\label{eq:LCFT.action}
S={1\over\pi} \int \partial \theta \; \bar\partial\bar\theta
\end{equation}

\noindent $\partial$ and $\bar\partial$ refer to the holomorphic
and antiholomorphic derivatives---that is, the derivatives
with respect to $z=x+iy$ and $\bar{z}=x-iy$.
$\theta$ and $\bar\theta$ are anticommuting
Grassmanian variables. The action has zero modes, which make
the partition function zero. If we normalize the action
by not integrating over the zero modes, we get
Wick contraction rules,
with each
contraction between $\theta (z_1)$ and
$\bar\theta (z_2)$ giving a factor of $-\log (z_1-z_2)$.

While boundary conformal field theories
are generally well 
understood~\cite{Cardy.mirror,Cardy.UnivCoeff},
boundary logarithmic conformal field theories 
possess a number of subtleties that are not well
understood.
Results on boundary LCFT are still to some degree
contradictory~\cite{LCFT.general,LCFT.general.2,BDY.LCFT.1,
BDY.LCFT.2,BDY.LCFT.3,BDY.LCFT.4}. However, basic
results from non-logarithmic boundary CFT
should still be expected to apply. In particular,
just as
for non-logarithmic boundary CFT's, 
as bulk operators are moved near a boundary, their
antiholomorphic pieces should behave like holomorphic pieces
at mirror locations across the 
boundary~\cite{Cardy.mirror,LCFT.general.2}.


\section{A PROOF THAT THE HEIGHT
VARIABLES HAVE DIFFERENT BULK FIELD IDENTIFICATIONS}
\label{sec:Simple.Proof}

The correlation functions in
\maybeeqs{\ref{eq:twoPoint.11}-\ref{eq:twoPoint.22}},
\maybeeqs{\ref{eq:f111}-\ref{eq:f122}}, and 
\maybeeqs{\ref{eq:defectnochange}-\ref{eq:defect3}}
show conclusively that the three height variables
are represented by different operators along
closed boundaries. As already discussed,
since boundary operators are derived from
OPE's of bulk
operators, this proves that the 
height variables
must be represented by different operators in the bulk as 
well~\cite{Cardy.mirror}. However,
it is worth noting that this conclusion can be
reached with a
simple argument,
based on general principles of conformal field theory,
without doing any detailed calculations.

Suppose that all four height variables 
were represented (up to multiplicative factors)
by the same field operator. 
The unit height variable is known, from its
two-point correlation, to have 
dimension two~\cite{Dhar.UnitCorrelations}, so,
by our assumption, all four height variables
would have scaling dimension two. 
The height probabilities get modified from their bulk
values, $p_{B,h}$ ($h=1,2,3,4$) near a boundary (closed or
open). 
Then one-point functions of
operators of dimension $d$ 
will decay as 
$1/y^d$, where
$y$ is the distance from the boundary,
and $d$ is the operator
dimension~\cite{Cardy.UnivCoeff,LCFT.general.2}. 

\begin{equation}
p_h(y)=p_{B,h}+{c_h \over y^2}+\dots\ ,
\end{equation}

\noindent for some constants $c_h$.
If the fields are normalized (to have zero expectation
value and coeficient -1 in two-point correlations),
then general CFT principles predict that the
coefficients of the $1/y^2$ terms
should be universal numbers, depending
only on the field and the boundary
condition~\cite{Cardy.UnivCoeff}; 
in particular, they should be independent of $h$.
So upon normalizing the height variables, the different
$c_h$ should all become $\tilde{c}$,
a number independent of $h$.
Since we are assuming
that all four height variables are represented by the
same field, the 1-1, 2-2, 3-3, and 4-4 correlations
should all have the same sign (negative), so 
this normalization should not change the signs of the
coefficients, and all the $c_h$'s should have
the same sign as $\tilde{c}$.
However, we need $\sum_{h=1}^4 c_h=0$, for
the four height probabilites to sum to one, so
the $c_h$ cannot all have the same sign. 
By contradiction, the four height variables must
be represented by different fields in the bulk.


\end{document}